\documentclass[%
 reprint,
 superscriptaddress,
 showpacs,
  nofootinbib,
 amsmath,amssymb,
 aps,
 prc,
]{revtex4-1}

\usepackage{verbatim}
\usepackage{graphicx}
\usepackage{epsfig}
\usepackage{amsfonts}
\usepackage{amsmath,amssymb}
\usepackage{bm}
\usepackage{hyperref}
\usepackage{color}

\begin{document}

\title{Nuclear three-body problem in the complex energy plane: Complex-Scaling-Slater method}

\author{A.T. Kruppa}
\affiliation{
Department of Physics and Astronomy, University of Tennessee, Knoxville, Tennessee 37996, USA
}%
\affiliation{Institute of Nuclear Research,
  P.O. Box 51, H-4001 Debrecen, Hungary}

\author{G. Papadimitriou}
\affiliation{
Department of Physics, University of Arizona, Tucson, AZ 85721, USA
}%
\affiliation{
Department of Physics and Astronomy, University of Tennessee, Knoxville, Tennessee 37996, USA
}%

\author{W. Nazarewicz}
\affiliation{
Department of Physics and Astronomy, University of Tennessee, Knoxville, Tennessee 37996, USA
}%
\affiliation{
Physics Division, Oak Ridge National Laboratory, Oak Ridge, Tennessee 37831, USA
}%
\affiliation{
Institute of Theoretical Physics, University of Warsaw, ul. Ho\.za 69,
PL-00-681 Warsaw, Poland }%

\author{N. Michel}
\affiliation{
National Superconducting Cyclotron Laboratory and Department of Physics and Astronomy, Michigan State University, East Lansing, Michigan 48824, USA
}%
\affiliation{
Department of Physics and Astronomy, University of Tennessee, Knoxville, Tennessee 37996, USA
}%


\begin{abstract}
{\bf Background:} The physics of open quantum systems is an interdisciplinary area of research. The nuclear ``openness"  manifests itself through  the presence of the many-body continuum representing various decay, scattering, and reaction channels.   As the radioactive nuclear beam experimentation  extends the known nuclear landscape towards the particle drip lines, the coupling to the  continuum space becomes exceedingly more important. Of particular interest are
weakly bound and unbound nuclear states appearing around particle thresholds.
Theories of such nuclei must take into account their open quantum nature.

{\bf Purpose:}  To describe open quantum systems, we introduce a Complex Scaling (CS) approach in the Slater basis. We benchmark it with the complex-energy  Gamow Shell Model (GSM) by studying energies and wave functions of the bound and unbound states of the two-neutron halo nucleus $^6$He viewed as an $\alpha+n+n$ cluster system.

{\bf Methods:} Both CS and GSM are applied to a translationally-invariant Hamiltonian with the two-body interaction approximated by the finite-range central Minnesota force. In the CS approach, we use
the Slater basis, which exhibits the correct asymptotic behavior at large distances. To extract particle densities from the back-rotated CS solutions, we apply the Tikhonov regularization procedure, which minimizes the ultraviolet numerical noise.

{\bf Results:} We show that the  CS-Slater method is both accurate and efficient. Its equivalence with  GSM has been demonstrated numerically for both energies and wave functions of $^6$He. One important technical aspect of our calculation was to fully retrieve the correct asymptotic behavior of 
a resonance state from the complex-scaled (square-integrable)
wave function.  While standard applications
of the inverse complex transformation to the complex-rotated solution provide  unstable results, the  stabilization method fully reproduces the GSM benchmark. We also propose a method to determine the smoothing parameter of the Tikhonov regularization.

{\bf Conclusions:} The combined suite of CS-Slater and GSM techniques has many attractive features when applied to nuclear problems involving weakly-bound and unbound states. While both methods can describe  energies, total widths,  and wave functions of nuclear states, the CS-Slater method -- if it can be applied -- can provide an additional information about partial energy widths associated with individual thresholds. 

\end{abstract}

\pacs{ 21.60.-n,21.60.Gx,21.60.Cs,21.10.Gv,02.60.-x}

\maketitle



\section{Introduction}

The physics of open quantum systems spans many areas of research, ranging from optical physics to nano science, to atomic, and to nuclear physics. Of particular interest are long-lived metastable states and broad resonances: they
carry rich information about  localized nucleonic states confined to the nuclear interior, about the multi-channel environment of scattering and decaying states, and about the coupling between these two spaces. With  exciting advances in  radioactive beam experimentation worldwide,  many  weakly-bound isotopes inhabiting the outskirts of the nuclear landscape can now be reached; they provide a fertile territory where
 to study generic properties of open quantum systems \cite{OpenQS}.

To develop a microscopic theoretical framework that would unify structural and reaction-theoretical aspects of the nuclear many-body system remains a challenge. A step in this direction is the unification of bound states and resonant phenomena, often enabled by high-performance computing, and there has been an excellent progress in this area
\cite{Gamow_Rmatrix,Nunes_Ian,Deltuva, Navratil1, *Navratil2,*Navratil3,Nollett,gaute_michel,ncgsm,gaute_Oxygen,*gaute_ca48}.

One possible strategy in this area is to relate the resonance parameters directly to the complex-energy eigenvalues of the effective Hamiltonian. To this end, one
can solve the many-body eigenproblem with  the hermitian Hamiltonian by imposing  specific  boundary conditions  \cite{review_GSM},
or one can construct a manifestly a non-hermitian effective  Hamiltonian \cite{Feshbach,Marek_rotter,Volya}. In both cases, the eigenstates that appear below the particle threshold are bound, and the complex-energy states above the threshold represent the many-body continuum.

The GSM \cite{review_GSM} and  CS  \cite{ykho,moiseyev,ikeda_review} methods
deal with effective non-hermitian Hamiltonians. In the GSM, one starts with a hermitian Hamiltonian and by imposing outgoing boundary conditions one ends up
with a complex-symmetric Hamiltonian matrix. 
In the CS method, a non-Hermitian Hamiltonian appears as a result of a complex rotation of coordinates. The corresponding  non-unitary transformation is  characterized by a real parameter $\vartheta$. The transformed  eigenstates 
 are square integrable;
this is a very attractive feature from the computational point of view.
Unfortunately, since the eigenvectors depend on $\vartheta$, they cannot be directly compared with
the eigenfunctions of the original Hamiltonian. To obtain the wave functions from the CS solutions,  the so called-back rotation
must be employed. Since in most cases the eigenproblem is solved numerically,
the back-rotation constitutes  an ill-posed inverse problem and a high-frequency ultraviolet 
noise appears \cite{backroterror,atkpal}.
We are aware of at least  two attempts \cite{pade,atombackrot} to overcome this problem. When the original wave function is
reconstructed by means of the Pad\'{e} approximation \cite{pade}, several calculations with different $\vartheta$ 
values can be carried out  to perform the analytical continuation. 
In Ref.~\cite{atombackrot}, special properties of the applied basis set were utilized to cure the
errors of the back rotated wave function. 
In this work, we will present a new approach to the problem of back-rotation. Our procedure does not depend on the type of  basis set used,
and it is based on sound mathematical foundations.  

The CS method has been successfully
applied in quantum chemistry to solve many-body problems with an extremely
high accuracy \cite{Bardsley,moi79,ykho,moiseyev,varga_positron} and also in nuclear physics, in  calculations of resonance parameters  \cite{atkrgm,kru99} and cluster systems \cite{ikeda_review,Aoyama95,*Aoyama95a,Myo01,katoalphad}. 
In the nuclear three body calculations, mainly Jacobi coordinates have been  employed.
In the cluster orbital shell model \cite{ikeda_review}, besides the ``V" type coordinate, also a ``T" type Jacobi coordinate has been used in order  to incorporate  correlations. 
In the field of quantum chemistry, on the other hand, mainly Hylleraas-type functions \cite{Hyll_basis,belen} are used, and the 
achieved accuracy for the helium atom  is spectacular \cite{drake1,drake2,korobov}.

In our CS calculations, we employ the Slater basis set \cite{Slater}, which is an approximation to the Hylleraas-type basis. The Slater 
wave functions  have  a correct asymptotic behavior, making them a perfect choice for the description of weakly-bound systems. 
A  basis set of similar type, the Coulomb-Sturmian functions,  has been  recently introduced  into the no-core shell model  framework \cite{vary}.
Those functions are in fact linear superpositions of Slater orbits.

In this work, the precision of the new CS-Slater method is tested against the results of the GSM calculations. For the sake of benchmarking, we consider the energies and wave functions of the
$0^+_1$ and $2^+_1$ states of $^6$He.
The paper is organized as follows. Section~\ref{Hami_Slater}  describes the Hamiltonian used, 
many-body methods, and configuration spaces employed. In Sec.~\ref{regularization} we discuss the difficulties related to the back-rotation  of the CS wave function and introduce the necessary regularization scheme. 
Section~\ref{results} presents the results  for $^6$He and the details of the CS-GSM benchmarking. Finally, conclusions and future plans are contained in  Sec.~\ref{concl}.

\section{Models and methods} \label{Hami_Slater}

\subsection{Three body Hamiltonian}
For the description of the ground and excited state of  $^6$He we  assume a cluster ($\alpha + n + n$) picture of the nucleus. Consequently,
we consider  a system of three particles with masses $m_i$ and single particle coordinates  $\bm{r}_{i}$, where $i=1, 2$ for neutrons and $i=3$ for the $\alpha$-core.
We introduce the relative coordinates $\bm{r}_{ij}=\bm{r}_{i}-\bm{r}_{j}$
and  $r_{ij}=|\bm{r}_{ij}|$.
The system Hamiltonian in the centre-of-mass frame reads:
\begin{eqnarray}\label{relham1}
H &=& -\frac{\hbar^2}{2\mu_1}\triangle_{\bm{r}_{13}}-\frac{\hbar^2}{2\mu_2}\triangle_{\bm{r}_{23}}
-\frac{\hbar^2}{m_3}\nabla_{\bm{r}_{13}}\nabla_{\bm{r}_{23}} \nonumber \\
&+& V_{12}(\bm{r}_{12})+ V_{13}(\bm{r}_{13})+V_{23}(\bm{r}_{23}),
\end{eqnarray}
where the reduced masses are:
\begin{equation}
\mu_1=\frac{m_1m_3}{m_1+m_3},~~~ \mu_2=\frac{m_2m_3}{m_2+m_3}.
\end{equation}
It is worth noting that the Hamiltonian 
\eqref{relham1} represents the intrinsic properties 
of the system, i.e., it is free from the spurious centre-of-mass motion. 
After introducing the single-neutron  Hamiltonian,
\begin{equation}\label{spham}
H_{i3}(\bm{r})= -\frac{\hbar^2}{2\mu_i}\triangle_{\bm{r}}+V_{i3}(\bm{r})~~~(i=1,2),
\end{equation}
the  Hamiltonian \eqref{relham1} can be written as:
\begin{equation}\label{relham2}
H = H_{13}(\bm{r}_{13})+H_{23}(\bm{r}_{23})+V_{12}(\bm{r}_{12})-\frac{\hbar^2}{m_3}\nabla_{\bm{r}_{13}}\nabla_{\bm{r}_{23}},
\end{equation}
where the last term represents a two-body recoil term, which originates from the transformation to the relative coordinate frame. 

\subsection{Complex Scaling Method}

The key element of the CS method is the complex-scaling operator 
$U(\vartheta)$, which transforms an arbitrary function $\chi(\bm{r}_{13},\bm{r}_{23})$ according to: 
\begin{equation} \label{complex_rot1}
U(\vartheta)\chi(\bm{r}_{13},\bm{r}_{23}) = 
e^{i 3\vartheta}\chi(  e^{i\vartheta}\bm{r}_{13}, e^{i\vartheta}\bm{r}_{23}).
\end{equation}
The transformed Shr\"{o}dinger equation becomes:
\begin{equation}\label{rot_Shroed}
H_{\vartheta}\Psi_{\vartheta} = E\Psi_{\vartheta}, 
\end{equation}
where 
\begin{equation}\label{rotated_H}
H_\vartheta = U(\vartheta)H U(\vartheta)^{-1}
\end{equation}
is a complex-scaled Hamiltonian:
\begin{eqnarray}\label{csham}
H_{\vartheta}\ &=&e^{-2i \vartheta}\left( -\frac{\hbar^2}{2\mu_1}\triangle_{\bm{r}_{13}}-\frac{\hbar^2}{2\mu_2}\triangle_{\bm{r}_{23}}
-\frac{\hbar^2}{m_3}\nabla_{\bm{r}_{13}}\nabla_{\bm{r}_{23}}\right) \nonumber \\
&+& V_{12}(e^{i\vartheta}\bm{r}_{12})+ V_{13}(e^{i\vartheta}\bm{r}_{13})+V_{23}(e^{i\vartheta}\bm{r}_{23}).
\end{eqnarray}
The exact eigenfunctions $\Psi(\bm{r}_{13},\bm{r}_{23})$ and $\Psi_\vartheta(\bm{r}_{13},\bm{r}_{23})$ of the Hamiltonians (\ref{relham1}) and (\ref{csham}) satisfy the following relation:
\begin{equation}\label{rot}
\Psi_\vartheta(\bm{r}_{13},\bm{r}_{23})= 
e^{i 3\vartheta}\Psi(  e^{i\vartheta}\bm{r}_{13}, e^{i\vartheta}\bm{r}_{23})
\end{equation}
or the so-called back rotation relation:
\begin{equation}\label{backrot}
\Psi(\bm{r}_{13},\bm{r}_{23})= 
e^{-i 3\vartheta}\Psi_\vartheta(  e^{-i\vartheta}\bm{r}_{13}, e^{-i\vartheta}\bm{r}_{23}).
\end{equation}

\begin{figure}[b] 
  \includegraphics[width=0.8\columnwidth]{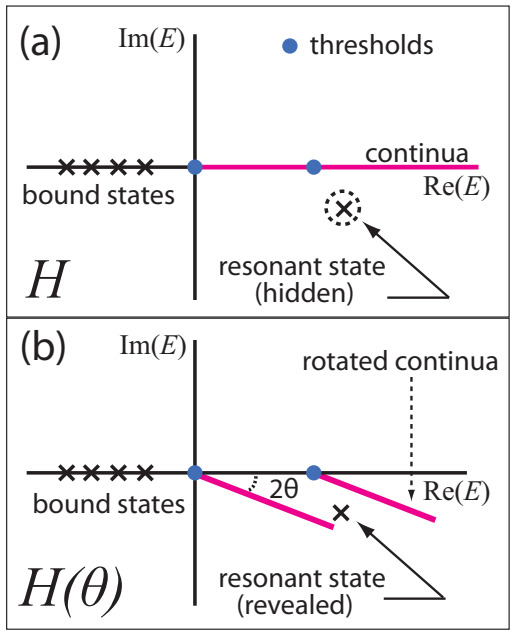}
  \caption[T]{\label{fig1}
  (Color online)  Illustration of the complex scaling transformation of a many-body Hamiltonian. Bound
states and many-body thresholds are invariant. Resonant eigenvalues correspond to poles of the resolvent or the $S$-matrix,
 are ``hidden" on a  sheet with $\vartheta$ = 0 (a), but are exposed if the cuts associated with many-body continua  are rotated (b) \cite{reinhardt}. }
\end{figure}

According to the Aguilar-Balslev-Combes theorem  \cite{abc1,abc2}, the resonant solutions of Eq.~\eqref{rot_Shroed}
are square integrable. This feature makes it possible to use   bound-state methods to solve  \eqref{rot_Shroed}, including
configuration interaction  \cite{ykho,moiseyev}, Faddeev and Faddeev-Yakubovsky \cite{lazauskas1,lazauskas2}, and
Coupled Cluster method \cite{CC_chem}. As illustrated in Fig.~\ref{fig1},
the spectrum of the rotated Hamiltonian (\ref{rotated_H}) consists of bound and unbound states.
The continuum part of the spectrum is represented by
cuts in the complex energy plane at an angle $2\vartheta$ with the real-energy
axis, originating at many-body thresholds.
The resonant spectrum consists of bound states lying on the negative real energy axis
and positive-energy resonances. One attractive feature of the CS method is that one does not need to apply directly any boundary condition to obtain the resonant states.
Through the CS transformation $U(\vartheta)$,
all resonant wave functions have decaying asymptotic behavior.
Even though the solution of the complex-rotated Hamiltonian $H_{\vartheta}$ is square integrable, the back-rotated wave function  is an outgoing solution of the Schr\"odinger
equation with the original Hamiltonian $H$. The back-rotation transformation, or analytical continuation, will be
investigated in the following. 

While the rotated non-resonant continuum states depend on the
rotation angle, resonant states should be independent of $\vartheta$.
In practical applications, however,  Eq.~\eqref{rot_Shroed} cannot be solved exactly and usually a truncated basis set is adopted. As a consequence,
the positions of resonant states  move slightly 
with  $\vartheta$ and/or the size of the (truncated) basis. Since
the dependence on $\vartheta$ is radically different for the continuum spectrum and the resonant states, there exist practical techniques to identify the resonance solutions.  One of them is the so-called $\vartheta$-trajectory method: using the generalization of the virial theorem to  complex energies, one finds that the  resonant
solution   must change little  with  $\vartheta$ around certain value of
$\vartheta=\vartheta_{\rm opt}$.   In this work, we checked carefully the dependence of  resonant states on both  $\vartheta$ and  basis parameters.

\subsubsection{Slater-basis expansion}

To solve the CS problem, we  use a finite Slater-type basis set \cite{Slater}.
Namely, the eigenstate of the original Hamiltonian is assumed to be
\begin{eqnarray}\label{wftot}
&\Psi^{JM}(\bm{x}_{13},\bm{x}_{23}) = \sum_{\{lj\}}\sum_{A} C_{A}^{\{lj\}} \chi_{A}^{\{lj\}}(r_{13},r_{23})\nonumber\\
&\times {\cal Y}_{\{lj\}}^{JMTT_z}(\bm{x}_{13},\bm{x}_{23}),
\end{eqnarray}
where the linear expansion coefficients $C_{A}^{\{lj\}}$ are determined by the Rayleigh-Ritz variational principle. Here $\bm{x}_{13}$, $\bm{x}_{23}$ denote the spatial  and spin-isospin coordinates of first and second particle, respectively. 
For brevity  we introduce the compact notation $\{lj\}=l_{13}, j_{13}, l_{23}, j_{23}$. 
Furthermore we introduce the spin-isospin part:
\begin{eqnarray}
&{\cal Y}_{\{lj\}}^{JMTT_z}(\bm{x}_{13},\bm{x}_{23})= 
\chi_{TT_z}(1,2) \times \nonumber\\
&\left [\left [ {\cal Y}_{l_{13}}(\bm{r}_{13})\otimes\chi_{1/2}(1)\right]^{j_{13}}
\otimes\left [ {\cal Y}_{l_{23}}(\bm{r}_{23})\otimes\chi_{1/2}(2)\right]^{j_{23}}\right ]^{JM}\nonumber,
\end{eqnarray}
where the  solid spherical harmonics are ${\cal Y}_{lm}(\bm{r})=r^l Y_{lm}(\hat{\bm{r}})$.
The symbol $[\otimes]^{JM}$ denotes the angular momentum coupling and $\hat{\bm{r}}_{ij}$ stands for
the  angular coordinates of $\bm{r}_{ij}$. The total isospin and single-nucleon spin  functions are, respectively,  denoted by $\chi_{T,T_z}(1,2)$ and  $\chi_{1/2}(i)\ i=1,2$.

For the radial part of the wave function we  use the product of Slater-type functions:
\begin{equation}\label{radform}
\chi_{A}^{\{lj\}}(r_{13},r_{23})=r_{13}^{n}e^{-\alpha r_{13}}\ r_{23}^{m} e^{-\beta r_{23}},
\end{equation}
where the non-linear parameters of the basis may depend on the quantum numbers $\{lj\}$ 
and they are denoted by $A=\{\alpha,n,\beta,m\}$.
At this point, we neglect the inter-nucleon distance $r_{12}$ in the radial part in order to span the same subspace of the Hilbert space as the GSM. 
(When the three-body wave function does not depend on the inter-particle distance $r_{12}$ one refers to the resulting set as the  Slater basis. If all three coordinates are considered, the basis set  is called Hylleraas basis.)
It has been found in quantum chemistry studies \cite{belen} that 
by neglecting  $r_{12}$ 
and by using  20-30 Slater orbits,  the total energy is extremely close to the results of full Configuration Interaction  calculations.

In the LS coupling, the wave function \eqref{wftot} can be written in the form:
\begin{eqnarray}\label{lswf}
&\Psi^{JM}(\bm{x}_{13},\bm{x}_{23}) = \sum_{\{lj\}}\sum_{LS}\sum_{A} C_{A}^{\{lj\}} \chi_{A}^{\{lj\}}(r_{13},r_{23}) \nonumber \\
&\times \gamma_{LS}(\{lj\})
 \left [{\cal Y}^{L}_{l_{13}l_{23}}(\bm{r}_{13},\bm{r}_{23})  \otimes\chi_S(1,2)\right ]^{JM} \nonumber\\
 &\times\chi_{TT_z}(1,2),
\end{eqnarray}
where
\begin{eqnarray}\label{caly}
&{\cal Y}^{LM}_{l_1l_2}(\bm{r}_1,\bm{r}_2)= ~~~~~~~~~~~~~\strut \nonumber \\
& \sum_{m_1,m_2}\langle l_1m_1,l_2m_2\vert LM\rangle
{\cal Y}_{l_1,m_1}(\bm{r}_1){\cal Y}_{l_2,m_2}(\bm{r}_2)
\end{eqnarray}
are the bipolar harmonics, $\chi_{SS_z}(1,2)$ are coupled total spin functions, and $\gamma_{LS}(\{lj\})$ are recoupling coefficients~\cite{law80}.
In the case of a many-body system, the trial wave function is expanded in a many-body antisymmetric basis in 
a coupled or uncoupled scheme. In our formalism, we use the fully antisymmetrized wave functions expressed in both  LS- and JJ-coupling schemes. 
The trial wave function of the CS Hamiltonian has the  same form as Eq.~(\ref{wftot}):
\begin{eqnarray}\label{wftotcs}
&\Psi_\vartheta^{JM}(\bm{x}_{13},\bm{x}_{23}) = \sum_{\{lj\}}\sum_{A} C_{A}^{\{lj\}}(\vartheta) \chi_{A}^{\{lj\}}(r_{13},r_{23})\nonumber\\
&\times {\cal Y}_{\{lj\}}^{JMTT_z}(\bm{x}_{13},\bm{x}_{23}), \nonumber \\
\end{eqnarray}
but the  expansion coefficients $C_{A}^{\{lj\}}(\vartheta)$ now depend on $\vartheta$ and 
they are determined using the  generalized variational principle.

\subsubsection{Two-body matrix elements in CS}

Since the CS wave function is of Slater type, one needs to develop a technique  to compute two-body matrix elements (TBMEs). In the following, we  shortly review a 
method developed in the context of atomic physics applications  \cite{efr73,efr86,dra78}.

Since we  employ the LS coupling scheme, for TBMEs  we  need to consider integrals of the type:
\small
\begin{eqnarray}\label{v12}
&\langle A'\{l'j'\} | V_{12} |A\{lj\} \rangle = \int d\tau\chi_{A'}^{\{l'j'\}}(r_{13},r_{23}) 
{\cal Y}^{L}_{l_{13}' l_{23}'}(\hat{\bm{r}}_{13},\hat{\bm{r}}_{23})^* \nonumber\\
&\times V_{12}(r_{12})\chi_{A}^{\{lj\}}(r_{13},r_{23})
{\cal Y}^L_{l_{13} l_{23}}(\hat{\bm{r}}_{13},\hat{\bm{r}}_{23}).
\end{eqnarray}
\normalsize
To compute (\ref{v12}), we make a coordinate transformation
to  the three scalar relative coordinates $r_{12}, r_{13}, r_{23}$ and three
Euler angles ($\Omega$) corresponding to 
a triangle formed by three particles. 
The volume element $d\tau=d\bm{r}_{13} d\bm{r}_{13}$ can be then written as 
$d\tau_rd\Omega$, where the radial volume element is given by 
$d\tau_r=dr_{12} dr_{13} dr_{23} \, r_{12} r_{13} r_{23}$, and $d\Omega$ corresponds to angular volume element involving the Euler angles. 
The angular integral
\begin{eqnarray}
& W_{l'_1l'_2,l_1l_2}^L(r_{12},r_{13},r_{23})= \nonumber \\
&\int d\Omega\  
{\cal Y}^{L}_{l'_1 l'_2}(\bm{r}_{13},\bm{r}_{23})^*
{\cal Y}^L_{l_1 l_2}(\bm{r}_{13},\bm{r}_{23})
\end{eqnarray}
can be calculated analytically \cite{dra78},  and the result is: 
\begin{eqnarray}\label{w1}
& W^L_{l'_1,l'_2,l_1,l_2}(r_{12},r_{13},r_{23}) = r_{13}^{l_1+l_1'}r_{23}^{l_2+l_2'} \times  \nonumber \\
& \sum_\lambda A(l'_1,l'_2,l_1,l_2,L,\lambda)  P_\lambda\left(\frac{r_{13}^2+r_{23}^2-r_{12}^2}{2r_{13}\,r_{23}}\right),
\end{eqnarray}
where
\begin{eqnarray}
&A(l'_1,l'_2,l_1,l_2,L,\lambda) = \frac{1}{2}(-1)^L \hat{l_1}\hat{l_2}\hat{l'_1}\hat{l'_2}(-1)^\lambda(2\lambda+1) \nonumber \\
&\times \left(
\begin{array}{ccc}
l'_1&l_1&\lambda\\
0&0&0
\end{array}
\right)
\left(
\begin{array}{ccc}
l'_2&l_2&\lambda\\
0&0&0
\end{array}
\right) 
 \left\{
\begin{array}{ccc}
l_1&l_2&L\\
l'_2&l'_1&\lambda
\end{array}
\right\},
\end{eqnarray}
with $\hat{j}\equiv \sqrt{2j+1}$.
The presence of the Legendre polynomial 
$P_\lambda$ in (\ref{w1})
shows that the function $W^L_{l_1,l_2,l'_1,l'_2}(r_{12},r_{13},r_{23})$ is a multinomial in the variables $r_{12},r_{13}$ and $r_{23}$.
The interaction matrix element (\ref{v12}), can now be written in a compact form:
\begin{eqnarray}\label{wig}
&\langle A'\{l'j'\} | V_{12} | A\{lj\} \rangle  = \\
&=\int_0^\infty dr_{13}\,r_{13} \int_0^\infty dr_{23}\, r_{23} 
\int_{\vert r_{13}-r_{23}\vert}^{r_{13}+r_{23}}dr_{12}\, r_{12} \nonumber \\
& \times \chi_{A'}^{\{l'j'\}}(r_{13},r_{23})\chi_{A}^{\{lj\}}(r_{13},r_{23}) \nonumber\\
& \times \, V_{12}(r_{12})W^L_{l'_{13},l'_{23},l_{13},l_{23}}(r_{12},r_{13},r_{23}).
\nonumber
\end{eqnarray}
Finally we determine the radial integrals.
Using the functional form of the basis (\ref{radform}) and the dependence of the function 
$W^L_{l_1,l_2,l'_1,l'_2}(r_{12},r_{13},r_{23})$  on the integration variables, 
it follows that the building block of the calculation is the integral:
\small
\begin{eqnarray}\label{radint}
&I^{(\lambda)}(n_{13},n_{23})=\int_0^\infty  dr_{13}\int_0^\infty dr_{23}
\int_{\vert r_{13}-r_{23}\vert}^{r_{13}+r_{23}} dr_{12}\ r_{12}
r_{13}^{n_{13}}r_{23}^{n_{23}}\nonumber\\
&\times V_{12}(r_{12})P_\lambda\left(\frac{r_{13}^2+r_{23}^2-r_{12}^2}{2r_{13}r_{23}}\right)
\exp(-a_{13}r_{13}-a_{23}r_{23}), 
\end{eqnarray}
\normalsize
where
\begin{equation}
a_{13}=\alpha'+\alpha, ~~~a_{23}=\beta'+\beta,
\end{equation}
and
\small
\begin{equation}
n_{13}=n'+l'_{13}+n+l_{13}+1, ~n_{23}=m'+l'_{23}+m+l_{23}+1.
\end{equation}
\normalsize
The integral (\ref{radint})  can be easily calculated if the form factor of the interaction is exponential, 
Yukawa-like,  or Coulomb  \cite{fro96}. 
For a Gaussian form factor (e.g.,  Minnesota force), the integral (\ref{radint}) is more involved and the relevant expressions are given in Appendix A.

\subsection{Gamow Shell Model}

\begin{figure}[t] 
  \includegraphics[width=0.8\columnwidth]{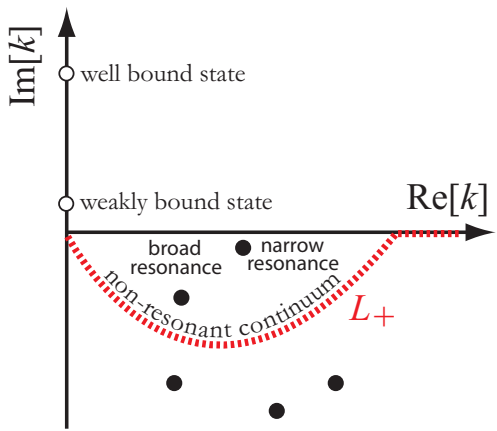}
  \caption[T]{\label{gsm_k_pic}
  (Color online) Berggren ensemble in the complex-$k$ plane used to generate the s.p. basis of the GSM.}
\end{figure}

The Gamow Shell Model is a complex-energy configuration interaction method \cite{review_GSM}, where the many-body Hamiltonian is
diagonalized in  a  one-body Berggren ensemble \cite{Berggren}
that contains both  resonant and non-resonant states. The total GSM wave function is expanded
in a set of basis states similar to Eq.~\eqref{wftot}. 
The basis functions $\psi^{(\alpha)}_{lj}(r)$  can here be represented by the eigenfunctions  of a single-particle (s.p.) Hamiltonian \eqref{spham} with
a finite-depth potential $V(r)$:
\begin{eqnarray}\label{gsm_sp_ham}
&\left ( -\frac{\hbar^2}{2\mu}\triangle_{\bm{r}} + V(r)\right)\psi^{(\alpha)}_{lj}(r)
\left [ Y_{l}(\hat{\bm{r}})\otimes\chi_{1/2}(1))\right]^{jm}\nonumber\\
&=\epsilon_\alpha\psi^{(\alpha)}_{lj}(r)\left [ Y_{l}({\hat{\bm{r}}})\otimes\chi_{1/2}(1))\right]^{jm}.
\end{eqnarray}
The resonant eigenstates (bound states and resonances), which correspond to the poles of the scattering $S$-matrix, 
are obtained by a numerical integration of the radial part of Eq.~\eqref{gsm_sp_ham}  assuming the outgoing boundary conditions:
\begin{equation}\label{boundary_cond}
        \psi(r) \stackrel{r \to 0}{=} r^{l+1}, ~~~~
        \psi(r) \stackrel{r \to \infty}{=} H^{+}_{l}(kr),
\end{equation}
where $H_{l}(kr)$ is a Hankel function (or Coulomb function  for protons).
The resulting
s.p. energies $\epsilon_{\alpha}$ and the associated linear momenta ($k_{\alpha} = \sqrt{2me_{\alpha}}/\hbar$) are in general complex.
As illustrated in Fig.~\ref{gsm_k_pic}, bound states are located on the imaginary momentum axis in the complex $k$-plane whereas the resonances are
located in its forth quadrant.
The s.p. Hamiltonian also generates  non-resonant states, which are solutions obeying scattering boundary conditions. The resonant and non-resonant states form a complete set (Berggren ensemble) \cite{Berggren,Lind,*Lind1}:
\begin{equation}\label{complet}
\sum_{b,r} | \psi_{b,r}^{\alpha} \rangle \langle \psi_{b,r}^{\alpha} | + \int_{L_{+}} dk |\psi_{k}^{\alpha} \rangle \langle \psi_{k}^{\alpha}| = 1, 
\end{equation} 
which is a s.p. basis of the GSM. In Eq.~(\ref{complet})
 $b$ (=bound) and $r$  (=resonance) are the resonant states, and the non-resonant states are distributed
along a complex contour $L_+$. In our implementations,
the continuum integral is discretized using a Gauss-Legendre quadrature.
The shape of the contour is arbitrary. The practical condition is that the contour should
enclose narrow resonances for a particular partial wave. Additionally, the contour is extended up to a certain momentum cut-off $k_{\rm max}$.
Then convergence of results is checked with respect to both the number of shells and the s.p. cut-off. For a sufficient number of points (shells), the basis (\ref{complet}) satisfies the 
completeness relation  to a very high accuracy.

The total wave function is expanded in the complete set of the Berggren's ensemble:
\small
\begin{eqnarray}\label{GSMwf}
&\Psi^{JM}(\bm{x}_{13},\bm{x}_{23}) = \sum_{\{lj\}}\sum_n  \sum_m C_{\{lj\}}^{(n,m)}\psi^{(n)}_{l_{13}j_{13}}(r_{13})\psi^{(m)}_{l_{23}j_{23}}(r_{23}) \nonumber\\
&\times {\cal Y}_{\{lj\}}^{JMTT_z}(\bm{x}_{13},\bm{x}_{23}).
\end{eqnarray}
\normalsize
Comparing Eqs.~\eqref{GSMwf} and \eqref{wftot}, we notice that the  GSM and CS-Slater wave functions differ by their  radial parts.
The  expansion coefficients $C^{(n)}_{lj}$'s are  determined variationally 
from  the eigenvalue problem:
\begin{equation}
\sum_{\alpha_1' \, \alpha_2'} \left( H_{\alpha_1 \alpha_2 \alpha_1'  \alpha_2'} - E C_{\alpha_1' \, \alpha_2'} \right) = 0,
\end{equation}
where,  $\alpha$ indices  represent the s.p.  $nlj$ quantum numbers.
Since the basis is in general complex, $H_{\alpha_1 \alpha_2 \alpha_1' \alpha_2'}$ is a non-Hermitian complex symmetric matrix.
The Berggren ensemble involves functions
which are not $L^2$-integrable. Consequently, normalization integrals and matrix elements of operators are 
calculated via the ``external" complex scaling technique \cite{Gya71}.

The GSM Hamiltonian is given by Eq.~\eqref{relham2}.
The s.p. potential $V(r)=V_{13}(r)=V_{23}(r)$  represents the interaction between the $\alpha$-core and the neutron, and
 $\mu=\mu_1=\mu_2$.
The same  interaction $V(r)$ is also used to generate the s.p. basis
\eqref{gsm_sp_ham}.

\subsubsection{Two-body matrix elements in GSM}

Once the basis is generated one needs to calculate TBMEs in the Berggren basis.
Since the Berggren basis is obtained numerically,
the standard  Brody-Moshinsky bracket technology \cite{Mosh1,*Mosh3,*Mosh2}, developed in the context of the harmonic oscillator (HO) s.p. basis, cannot be employed.
To overcome this difficulty,  we expand the NN interaction in a truncated HO basis \cite{hagen_morten_michel}:
\begin{equation}
V_{NN} = \sum_{\alpha \beta \gamma \delta}^{N_{\rm max}} |\alpha \beta \rangle \langle \alpha \beta|V_{NN}| \gamma \delta \rangle \langle \gamma \delta |.
\label{HO_exp}
\end{equation}
The TBMEs in the Berggren ensemble are given by:
\begin{equation}\label{inter_ho_exp}
\langle \widetilde{ab}| V_{NN} | cd \rangle = \sum_{\alpha \beta \gamma \delta}^{n_{\rm max}} \langle \widetilde{ab}|\alpha \beta  \rangle \langle \alpha \beta|V_{NN}| \gamma \delta \rangle \langle \gamma \delta | cd \rangle,
\end{equation}
where the Latin letters denote Berggren s.p. wave functions  and Greek letters -- HO states. 
Due to the Gaussian fall-off of  HO states, no external complex scaling is needed for the calculation of  the overlaps $\langle \alpha \beta | a b \rangle$. Moreover, matrix elements $\langle \alpha \beta|V_{NN}| \gamma \delta \rangle$ of the NN interaction in the HO basis  can be conveniently calculated using the Brody-Moshinsky technique \cite{Mosh1,*Mosh3,*Mosh2}. 
This method of treating the TBMEs of the interaction is  similar to the technique based on a separable expansion of the potential \cite{gyar_kruppa}.
The HO basis depends  on the oscillator length $b$, which is an additional parameter. However, as it was demonstrated in Refs.~\cite{hagen_morten_michel,Mic10}, GSM eigenvalues and eigenfunctions converge for a sufficient number of $n_{\rm max}$, and
the dependence of the results on $b$ is negligible. We shall return to this point  in Sec.~\ref{res:energies} below.

\subsubsection{Model space of GSM}

The CS and GSM calculations for the 0$^+$ g.s. of $^6$He have been performed in a model space of four partial waves: $p_{3/2}$, $p_{1/2}$, $s_{1/2}$, and $d_{5/2}$.
The Berggren basis consists of the 0$p_{3/2} $ resonant state, which is found at an energy of $0.737 -i0.292$\,MeV, and the $p_{3/2}$ complex contour in order to satisfy the Berggren's completeness relation.  The remaining partial waves 
$p_{1/2}$, $s_{1/2}$, and $d_{5/2}$ are taken along the real axis. 
Each contour is discretized with sixty points; hence, our one-body space consists of 241 neutron shells total. Within such a basis, results are independent on the contour extension in the
$k$-space. For the present calculation we used  a $k_{\rm max} = 3.5$\,fm$^{-1}$. The finite range Minnesota interaction was expanded in a set of HO states.
For the g.s., when a relatively large set of HO quanta is used, the dependence of the results on the HO parameter $b$ is negligible. 
We took $b = 2$\,fm and we used all HO states with up to $n_{\rm max} = 18$ radial nodes. Since the $s$ wave
enters the Berggren ensemble, in order to satisfy the Pauli
principle between core and valence particles we project out the 
Pauli forbidden  $0s_{1/2}$  state ($b=1.4$\,fm) using the
Saito orthogonality-condition model \cite{Saito}.

For the excited unbound 2$^+$ state of $^6$He we limit ourselves to a $p_{3/2}$ model space. As concluded in Ref.~\cite{gsm_radii}, the structure of this state is dominated by a $(p_{3/2})^2$ parentage. Moreover, in this truncated
space the neutron radial density becomes less localized since the 2$^+$
becomes less bound when the model space is increased. 
The width of this state increases from $\sim$250\,keV in the ($p_{3/2}$, $p_{1/2}$, $s_{1/2}$,  $d_{5/2}$) space to  $\sim$580\,keV  in the truncated space of $p_{3/2}$ waves. Dealing with a broader resonance facilitates benchmarking with CS back-rotation results and helps pinning down dependence on HO parameters in GSM calculations. The $p_{3/2}$ continuum was discretized with a maximum of 60 points.  This ensures fully converged results with respect
to the Berggren basis (both the number of discretization points and  $k_{\rm max}$).

\section{Back rotation: from Complex Scaling to Gamow states}\label{regularization}

Even if the  energies of resonant states in CS and GSM are the same, the wave functions 
are different (see Eqs. \eqref{rot} and \eqref{backrot}). This implies that the respective expectation values of an observable 
$\hat O$ in states $\Psi(\bm{r}_{13},\bm{r}_{23})$ and $\Psi_\vartheta(\bm{r}_{13},\bm{r}_{23})$ cannot be compared directly. Moreover, when the wave function $\Psi_\vartheta(\bm{r}_{13},\bm{r}_{23})$ 
is used,  one has  to deal with  the  transformed operator: 
\begin{equation}\label{rot_operator}
\hat O_\vartheta = U(\vartheta)\hat O U(\vartheta)^{-1}.
\end{equation}
In some cases, it is straightforward  to derive the transformed operator. For instance, 
in the calculation of the root-mean-square radius, the transformed operator is  
$e^{2i\vartheta}\bm{r}_{13}^2+e^{2i\vartheta}\bm{r}_{23}^2$. The transformed recoil operator is given by $-e^{-2i\vartheta}\frac{\hbar^2}{m_3}\nabla_{\bm{r}_{13}}\nabla_{\bm{r}_{23}}$, and the angular correlation function is the mean value of the operator $\delta(\theta_{12}-\theta)$, where
$\theta_{12}$ is the angle between the vectors $\bm{r}_{13}$ and $\bm{r}_{23}$. 
For the radial density, the situation is not that simple and we shall discuss this point  in the following.

In order to retrieve the Gamow wave function of the original Schr\"{o}dinger equation, it is tempting to carry out a direct back-rotation of the CS wave function (\ref{wftot}):
\begin{eqnarray}\label{wfbackrot}
& e^{-i3\vartheta}\sum_{\{lj\}}\sum_{A} C_{A}^{\{lj\}}(\vartheta) \chi_{A}^{\{lj\}}(e^{-i\vartheta}r_{13},e^{-i\vartheta}r_{23})\nonumber\\
& \times {\cal Y}_{\{lj\}}^{JMTT_z}(\bm{x}_{13},\bm{x}_{23}) .
\end{eqnarray}
It turns out, however, that this method is numerically unstable.  Even for one particle moving in a potential well, 
 the direct back-rotation leads to unphysical large oscillations in  the wave 
function \cite{atkpal,backroterror}. To prevent this, a proper regularization procedure needs to be applied \cite {chu08,chu09}.

The radial density is defined as the mean value of the operator:
\begin{equation}\label{denop}
\frac{1}{2}\left[\delta(r_{13}-r)+\delta(r_{23}-r)\right].
\end{equation}
Using the CS wave function (\ref{wfbackrot}) and the Slater-type radial basis functions 
(\ref{radform}), the density can be casted into the form:
\begin{equation}
\rho_\vartheta(r)=r^2\sum_j C_j(\vartheta) r^{n_j} \exp(-a_j r),
\end{equation}
where C$_j(\vartheta)$ are related to the linear expansion parameters \eqref{wftotcs}, 
obtained from the diagonalization
of the complex scaled Hamiltonian \eqref{rot_Shroed}. 
If we consider the direct back-rotated wave function, the radial density is given by:
\begin{equation}\label{fdefcs}
\rho^{\rm back}_\vartheta(r)=e^{-i\vartheta}\tilde\rho_\vartheta(e^{-i\vartheta}r),
\end{equation}
where
\begin{equation}\label{fdef}
\tilde\rho_\vartheta(r)=r^2\sum_j C_j(\vartheta) r^{n_j} \exp(-a_j r).
\end{equation}
The   factor $r^2$ comes from the volume element when the Dirac-delta function in 
(\ref{denop}) is integrated. 
We shall see that the density
calculated in this way leads to extremely inaccurate results. In the following, we briefly show how to calculate the density of the original Gamow state using the CS wave function. Illustrative numerical examples will be presented in  Sec.~\ref{density_tikhonov}.

We may consider Eq. (\ref{fdef}) as a definition of a function defined along the non negative real axis 
and $\tilde\rho_\vartheta(e^{-i\vartheta}r)$ can be viewed as an attempt to extend (\ref{fdef}) into the complex plane. However, since the coefficients  $C_i(\vartheta)$  obtained  numerically are not accurate enough, and moreover the Slater expansion is always truncated,   the analytical continuation 
of $\tilde\rho_\vartheta$ is not a simple task.  To find a stable solution, we  apply a method
based on the theory of Fourier transformations.
We first extend $\tilde\rho_\vartheta(r)$ from $(0,\infty)$ to $(-\infty,\infty)$ by means of the mapping: 
\begin{equation}\label{fdeftrans}
f_\vartheta(x)=\tilde\rho_\vartheta(r_0 e^{-x}).
\end{equation}
The Fourier transform of (\ref{fdeftrans}) is:
\begin{eqnarray}\label{fourier1}
&\hat f_\vartheta(\xi)=\frac{1}{\sqrt{2\pi}}\int_{-\infty}^\infty e^{-i x \xi}f_\vartheta (x) \,dx = \nonumber \\
&=\frac{1}{\sqrt{2\pi}}\sum_j C_j(\vartheta) r_0^{n_j+2}\frac{\Gamma(n_j+2+i\xi)}{(r_0 a_j)^{n_j+2+i\xi}},
\end{eqnarray}
where $\xi$ and $x$ are dimensionless variables.

Usually,  $\hat f_\vartheta$  is determined with an error, which results in the appearance of high-frequency oscillations in $f_\vartheta$. 
Now we shall apply the Tikhonov smoothing \cite{Tikh_orig} to $f_\vartheta (x+iy)$. To this end, we perform
the analytical continuation of $f_\vartheta(x)$  to 
the complex plane $x+iy$  \cite{chu08}:
\begin{equation}\label{fourier_analytic_cont}
f_\vartheta(x+iy)=\frac{1}{\sqrt{2\pi}}\int_{-\infty}^\infty d\xi\, e^{-i (x+iy) \xi}\hat f_\vartheta (\xi).
\end{equation}
The Tikhonov regularization \cite{chu09} removes the ultraviolet noise in \eqref{fourier_analytic_cont} by introducing a smoothing function:
\begin{eqnarray}\label{tikh_formula}
f^{reg}_\vartheta(x+iy) &=& \frac{1}{\sqrt{2\pi}}\int_{-\infty}^\infty e^{-i (x+iy) \xi} \nonumber \\
&\times& \frac{\hat f_\vartheta (\xi)}{1+\kappa e^{-2y\xi}} d\xi,
\end{eqnarray}
where $\kappa$ is the Tikhonov smoothing parameter. In the actual calculation we take  $x=-\ln(r/r_0)$, $y=\vartheta$, and $r_0=1$\,fm.

\section{Results}\label{results}

For the neutron-core interaction we employ the KKNN potential \cite{KKNN} and the interaction between the valence neutrons is approximated by the Minnesota force \cite{LeMere_Tang}.
We study the convergence properties of the CS-Slater method not only for energies of $0^+_1$ and $2^+_1$ of  $^6$He and  individual energy components, but also for radial properties and spatial correlations.

\subsection{Energies}\label{res:energies}

According to \eqref{relham2} the total Hamiltonian of $^6$He is the sum of one-body terms $H_{13}(\bm{r}_{13})+H_{23}(\bm{r}_{23})$ and two-body terms
$-\frac{\hbar^2}{m_3}\nabla_{\bm{r}_{13}}\nabla_{\bm{r}_{23}}+V_{12}(\bm{r}_{12})$. 
\begin{figure}[htb]
  \includegraphics[width=0.8\columnwidth]{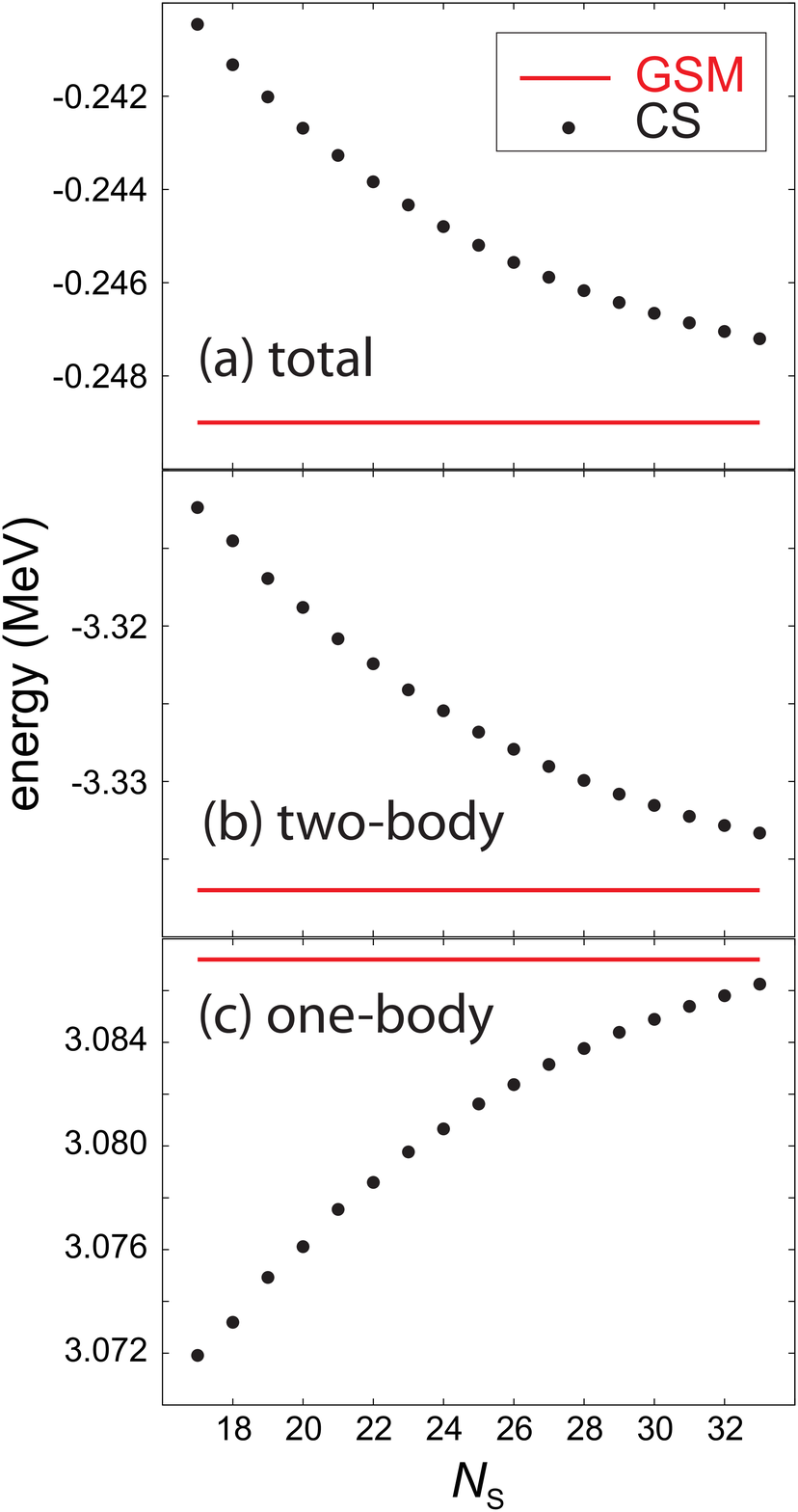} 
  \caption[T]{\label{conv_Nmax}
  (Color online) Convergence of the $^6$He total g.s. energy, two-body, and one-body terms, with respect to the number of 
  Slater orbitals $N_{\rm S}$ for $\alpha=\beta=0.8$.}
\end{figure}
Figure~\ref{conv_Nmax} illustrates the   convergence of the CS energies with respect to the basis size $N_{\rm S} \ge n+m$ (see Eq.~\eqref{radform} for notation). A similar type of restriction was used in Refs.~\cite{drake1,drake2} in order to avoid the linear dependence of the basis functions. For the non-linear parameters of the Slater basis we assumed the value $\alpha=\beta=0.8$.
The dependence on the  Slater basis parameter $\alpha$ is shown in 
Fig.\ref{non_lin_dep} for $N_{\rm S}=27$.

In Figs.~\ref{conv_Nmax} and \ref{non_lin_dep}, horizontal solid lines correspond to GSM results. The maximum difference between CS and GSM energies is of the order of 2 keV for the total energy,
two-body, and one-body terms. As can be seen in Fig.~\ref{non_lin_dep},  two-body and one-body terms have no minima with respect to  $\alpha$. This is expected as it is the total energy that that is supposed to exhibit a variational minimum, not its individual contributions.
\begin{figure}[h!] 
  \includegraphics[width=0.8\columnwidth]{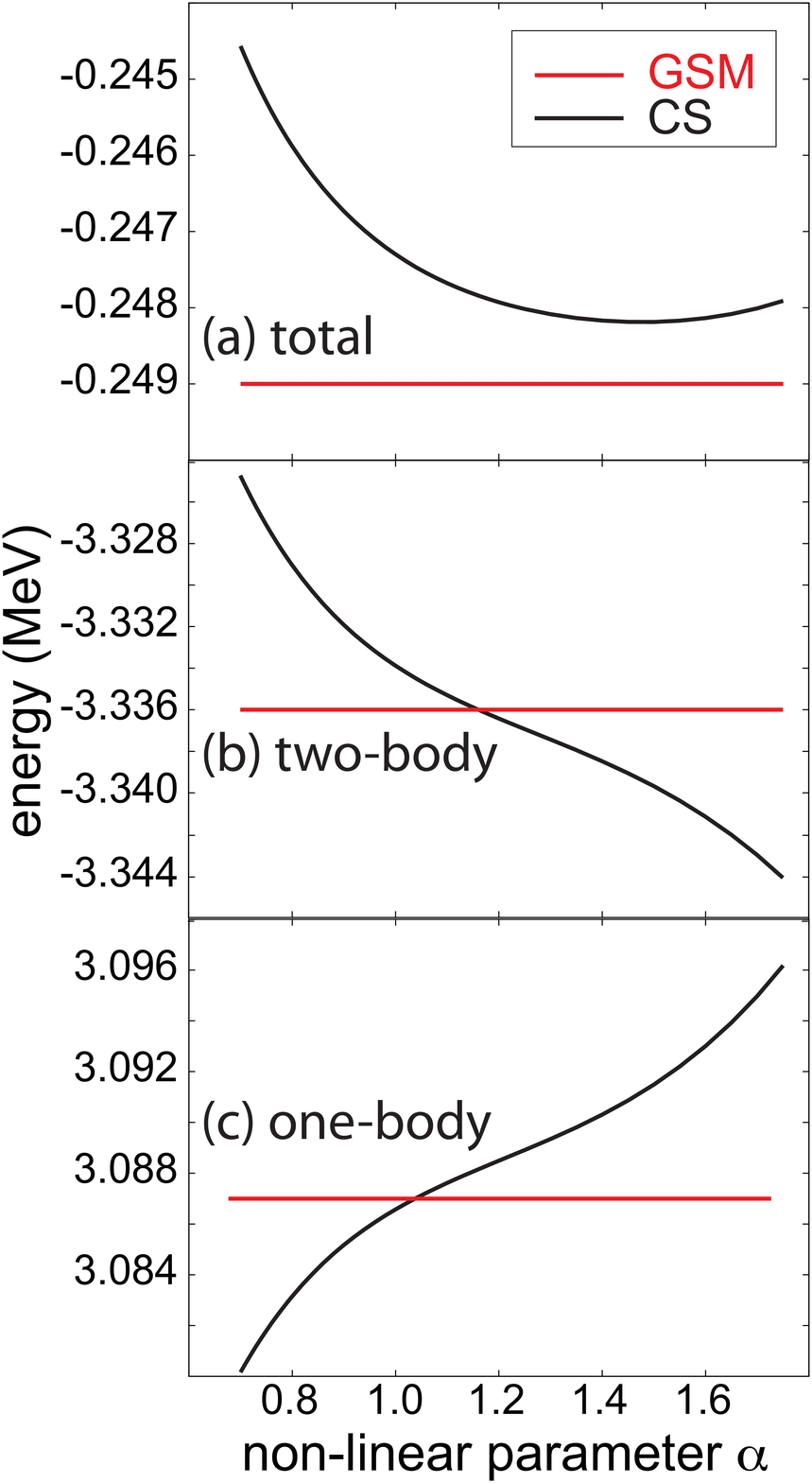}
  \caption[T]{\label{non_lin_dep}
  (Color online) Similar as in Fig.~\ref{conv_Nmax} but versus
   the non-linear Slater basis parameter $\alpha=\beta$
  for $N_{\rm S}=27$.}
\end{figure}
The two and one body terms coincide with the GSM result for a slightly different variational parameter ($\alpha \sim 1.1$) than the one that corresponds to the 
minimum of the total energy ($\alpha = 1.5$). Nevertheless, the difference at the minimum is very small, of the order of 2 keV.

Table \ref{Tab:1} displays the energy budget for the bound g.s. configuration of $^6$He in GSM and  CS methods. Even though it is not necessary to use CS 
for  a bound state, we also show values for $\vartheta$ = 0.2, for the reasons that will be explained later in Sec.~\ref{density_tikhonov}. 
In this case, the expectation value of the transformed operator $\hat O_\vartheta = U(\vartheta)\hat O U(\vartheta)^{-1}$ was computed.
It is seen that the excellent agreement is obtained between GSM and both CS variants not only for the total energy but also for {\em all} Hamiltonian terms.
\begin{table}[ht]
\caption{\label{Tab:1} Energy decomposition of $^6$He g.s. Values are in MeV. }
 \begin{ruledtabular}
\begin{tabular}{lrrrr}
$\langle \hat{O} \rangle$ & GSM~\strut & CS ($\vartheta = 0$) & CS ($\vartheta = 0.2$)~~~~~\strut \\
\hline
$\langle \, \hat{H} \rangle$ & $-$0.249 & $-$0.24\textcolor{blue}{7}~~~\strut & $-0.24\textcolor{blue}{7} + i1.1\times 10^{-3}$ \\
$\langle \, \hat{T} \rangle$ & 24.729 & 24.7\textcolor{blue}{31}~~~\strut & $24.7\textcolor{blue}{33} - i7.27\times 10^{-3}$ \\
$\langle \, V_{c-n} \rangle$ & $-$21.642 & $-$21.64\textcolor{blue}{5}~~~\strut & $-21.64\textcolor{blue}{7} + i4.76\times 10^{-3}$  \\
$\langle \, V_{nn} \rangle$ & $-$2.711 & $-$2.71\textcolor{blue}{0}~~~\strut  &  $-2.71\textcolor{blue}{0} +  i3.11\times 10^{-3}$ \\
$\langle \, \frac{ \vec{p_{1}} \cdot \vec{p_2}}{m_3} \rangle$ & $-$0.625 & $-$0.62\textcolor{blue}{3}~~~\strut & $-0.62\textcolor{blue}{3} + i5.04\times 10^{-3}$ 
\end{tabular}
\end{ruledtabular}
 \end{table}

We now move on to the 2$^{+}$ unbound excited state of $^6$He. 
To assess the accuracy of computing this state in GSM, we test the sensitivity of calculations to the HO expansion  \eqref{inter_ho_exp}. It is worth noting  that in the GSM only the two-body interaction and recoil term  are treated within the HO expansion.
The kinetic term is calculated in the full Berggren basis; hence,  the system maintains the correct asymptotic behavior. Moreover, for the $2^+$  state 
in the  $p_{3/2}$ model space, the recoil term vanishes.
\begin{figure}[h!] 
  \includegraphics[width=0.8\columnwidth]{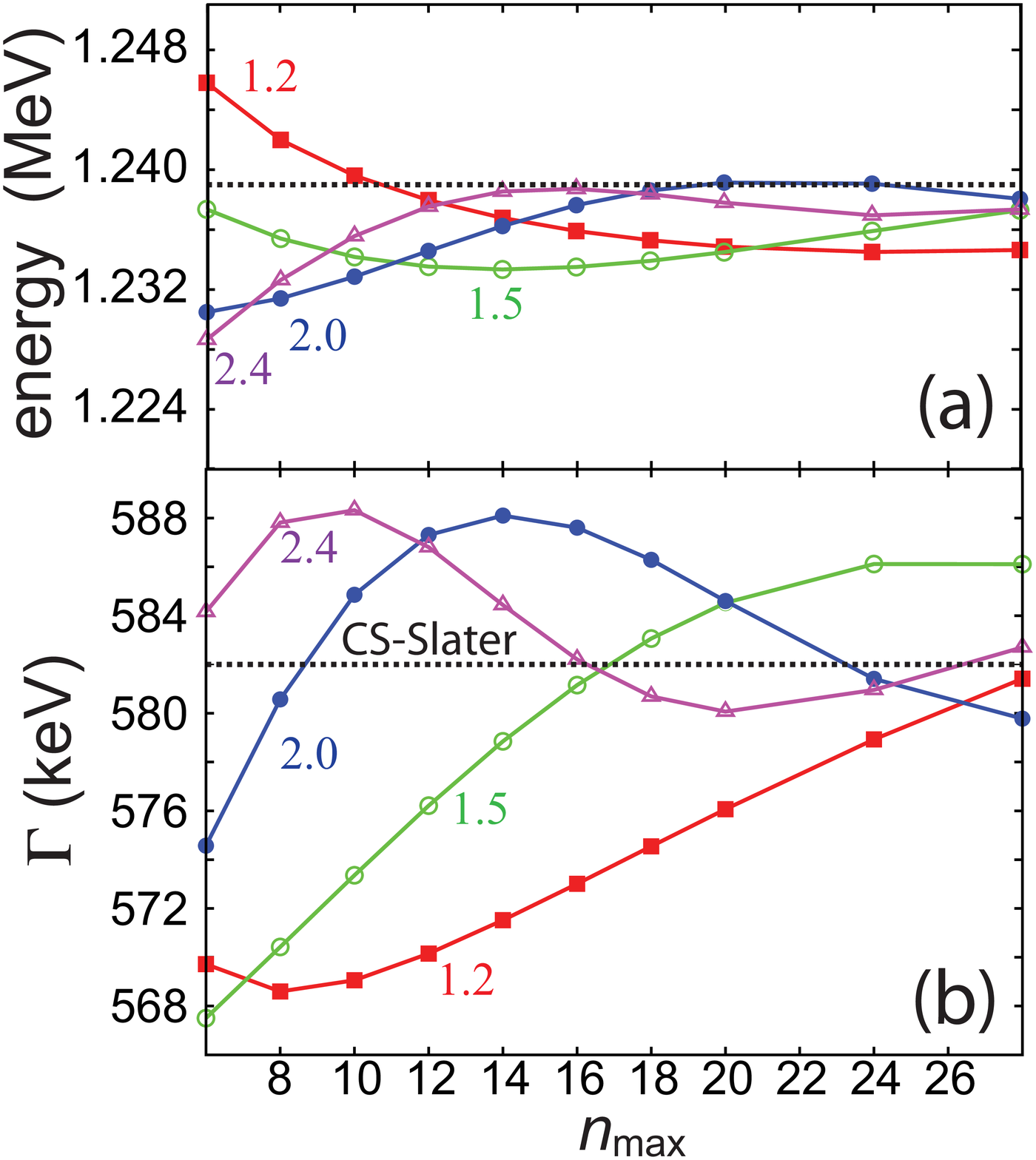}
  \caption[T]{\label{nmax_dependence}
  (Color online) Dependence of the  energy (a) and width (b)  of the unbound 2$^+_1$  state in $^6$He calculated with GSM on the HO expansion parameters  $n_{\rm max}$ and $b$ (=1.2, 1.5, 2.0, and 2.4\,fm) in Eq.~\eqref{inter_ho_exp}. The CS-Slater result is marked by a dotted line.}
\end{figure}
The resonance position in the CS-Slater method is determined with the  $\vartheta$-trajectory method.
Figure~\ref{nmax_dependence} displays the result of our tests. 
Overall, we see a  weak dependence of the energy and width of the  2$^+$ state predicted in GSM on the HO expansion parameters $n_{\rm max}$ and $b$. The increase of $n_{\rm max}$
 from  6 to 28  results in energy (width) change of $\sim$20\,keV ($\sim$10\,keV). With increasing $n_{\rm max}$, 
the results become less dependent on the
oscillator length $b$. 
For the real part of the energy, there appears some stabilization at large values of $n_{\rm max}$. 
but the pattern is different for different values of $b$. 
The most stable results are obtained with $b=2$\,fm, where we find a broad  plateau for both the  energy and the energy modulus \cite{Moi78,moiseyev,Rot09}  for $n_{\rm max}>16$. We adopt the value of $b_{\rm opt}=2$\,fm for the purpose of further benchmarking.

The pattern for the width is similar, with no clear
plateau but very small differences at large $n_{\rm max}$. Such a behavior is not unexpected. While the variational arguments do not apply to the interaction but to the trial wave function \cite{Moi78,moiseyev,Rot09}, one
can demonstrate \cite{hagen_morten_michel,Mic10} that while the matrix elements exhibit weak converge
with $n_{\rm max}$, eigenvectors and energies show strong convergence. However, the actual convergence is very slow for broad resonant states.

Based on our tests presented in Fig.~\ref{nmax_dependence} we conclude that the numerical error of GSM, due to HO expansion, on the energy and width of the $2^+_1$ resonance in $^6$He is $\sim 2$\,keV. This accuracy is more than needed  to carry out the CS-GSM benchmarking.

Table \ref{Tab:2} displays the energy budget  for the unbound 2$^+_1$ state of $^6$He.
\begin{table}[ht]
\caption{\label{Tab:2} Similar to Table~\ref{Tab:1} but for the   2$^+_1$  resonance.
In GSM calculations, we used $b_{\rm opt}=2$\,fm  and $n_{\rm max}=20$ (GSM$_{\rm I}$)
and $n_{\rm max}=24$ (GSM$_{\rm II}$). The optimal scaling angle $\vartheta_{\rm{\rm opt}}=0.43$ was obtained with the $\vartheta$-trajectory method.
}
 \begin{ruledtabular}
\begin{tabular}{crrrr}
$\langle \hat{O} \rangle$ & CS ($\vartheta$ = $\vartheta_{\rm{\rm opt}}$) & GSM$_{\rm I}$~~~~~\strut  & GSM$_{\rm II}$~~~~~\strut  \\  
\hline
$\langle \, \hat{H} \rangle$ & $1.239 - i0.291$ & $1.239 - i0.29\textcolor{blue}{2}$ & $1.239 - i0.29\textcolor{blue}{0}$ \\
$\langle \, \hat{T} \rangle$ & $17.340 - i7.949$ & $17.3\textcolor{blue}{11} - i7.\textcolor{blue}{825}$ & $17.\textcolor{blue}{221} - i7.\textcolor{blue}{766}$ \\
$\langle \, V_{c-n} \rangle$ & $-15.831  + i7.408$ & $-15.8\textcolor{blue}{05} + i7.\textcolor{blue}{288}$ & $-15.\textcolor{blue}{717} + i7.\textcolor{blue}{231}$ \\
$\langle \, V_{nn} \rangle$ & $-0.270  + i0.250$ & $-0.2\textcolor{blue}{67} + i0.2\textcolor{blue}{44}$ & $-0.2\textcolor{blue}{65} + i0.2\textcolor{blue}{44}$
\end{tabular}
\end{ruledtabular}

\end{table}%
We show two variants of  GSM calculations in which  the interaction was expanded in a HO basis with $b_{\rm opt} = 2$\,fm and $n_{\rm max}$ = 20 (GSM$_{\rm I}$) and 24
(GSM$_{\rm II}$).
The real parts of the total  energy are identical in both methods  up to the third digit, and the imaginary parts up to second digit.
For the other parts of the Hamiltonian, results are not as precise as for the g.s. calculations in Table \ref{Tab:1}; nevertheless, we obtain an overall satisfactory agreement. It is encouraging, however,
that for the total complex energy the agreement is excellent. The benchmarking results presented in this section demonstrate the equivalence of GSM and CS-Slater methods for energies of bound and unbound resonance states. In the following, we shall see that this equivalence also holds for the many-body wave functions.

\subsection{One-body densities} \label{density_tikhonov}

\begin{figure}[h!] 
  \includegraphics[width=0.8\columnwidth]{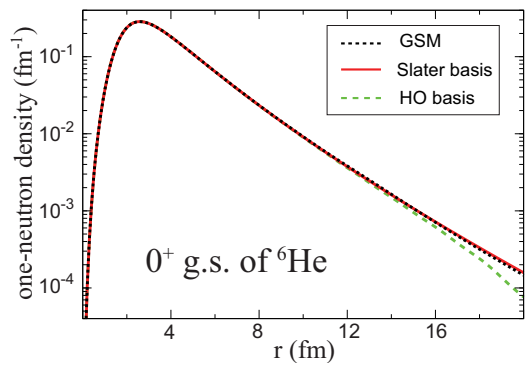}
  \caption[T]{\label{den_gsm_cs_ho}
  (Color online) Ground-state one-neutron radial density in $^6$He predicted with  GSM, CS-Slater, and HO basis sets.}
\end{figure}

To assess the quality of wave functions calculated with GSM and CS-Slater, we first calculate the radial one-neutron density of the g.s. of $^6$He.
Figure~\ref{den_gsm_cs_ho} shows that both methods  are consistent with each other and they correctly predict exponential fall-off at large distances. 
We also display the one-neutron density obtained with the radial part
of the wave function \eqref{wftot}  spanned by the radial HO basis states with
$b=2$\,fm and $n_{\rm max}=18$. As expected, the HO result falls off too quickly at very large distances due to the incorrect asymptotic behavior.

The g.s. of $^6$He is a bound state; hence,  its description does not require a complex rotation of the Hamiltonian. Nevertheless, it is instructive to study  the effect of CS on its radial properties.
\begin{figure}[h!] 
  \includegraphics[width=\columnwidth]{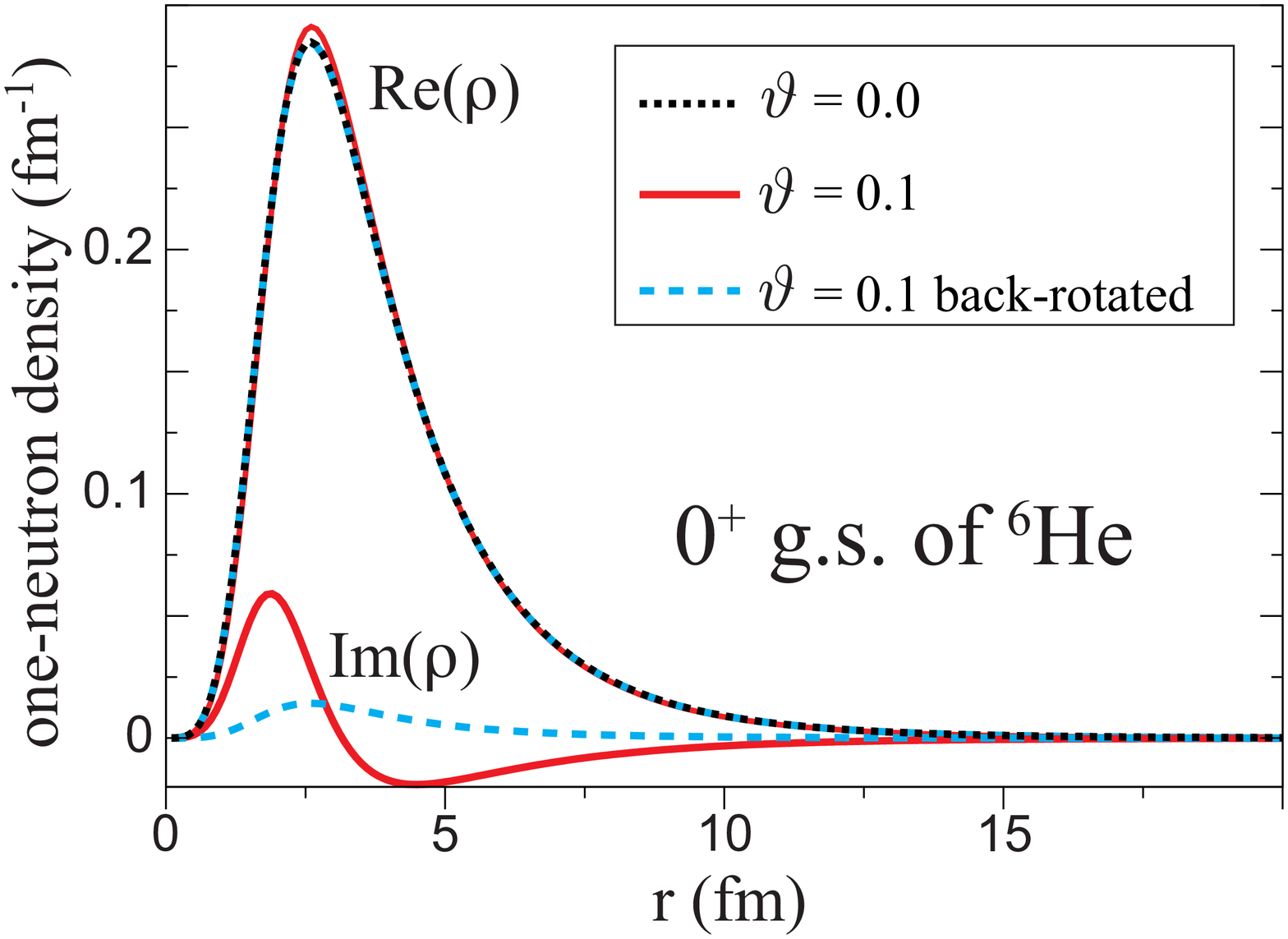}
  \caption[T]{\label{den_theta}
  (Color online) Ground-state one-neutron radial density in $^6$He predicted in CS-Slater using $\vartheta=0$ (dotted line) and 0.1 (solid line). The back-rotated $\vartheta=0.1$ result is marked by a dashed line.
  }
\end{figure}
Figure~\ref{den_theta}  shows  the g.s. one-neutron density obtained in CS-Slater using  $\vartheta=0.1$.    
For comparison we also display the
unscaled ($\vartheta=0$) density of Fig.~\ref{den_gsm_cs_ho}.
We see that the one-particle density is $\vartheta$-dependent and 
for $\vartheta>0$  it acquires  an imaginary part. 
Since the integral of the density is normalized to 1,  the integral 
of the imaginary part should be zero. This  
was checked numerically to be indeed the case. Since the  back-rotated density should be equivalent to the unscaled or GSM one, its imaginary part should vanish. However, as seen in Fig.~\ref{den_theta}, the back-rotated density at $\vartheta=0.1$ is nonzero. This is indicative of serious problems with 
back-rotation in CS, if this method is applied directly \cite{backroterror,atkpal}.

In order to investigate back-rotation in more detail, we consider the  $2^+_1$ resonance in $^6$He. As  in Sec.~\ref{res:energies}, we limit ourselves to a $p_{3/2}$ model space to better see the effect of back-rotation; by adding more partial waves, the $2^+$ state becomes more localized and the CS density resembles the GSM result.
\begin{figure}[h!] 
  \includegraphics[width=\columnwidth]{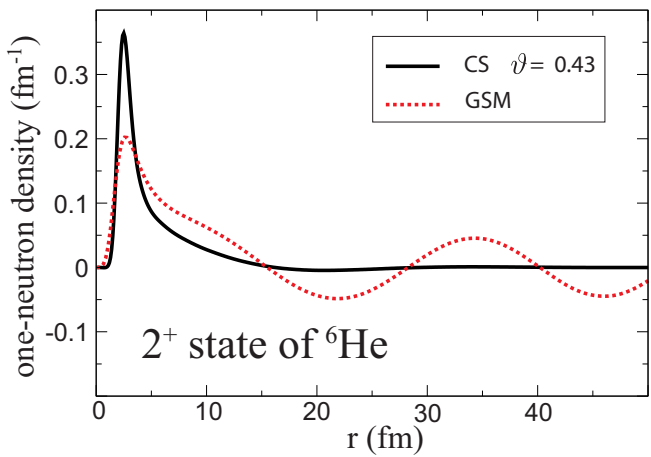}
  \caption[T]{\label{GSM_vs_L2_state}
  (Color online) Real part of one-neutron radial density for the unbound  2$^+$ state in $^6$He obtained in GSM (solid line)  and CS-Slater ($\vartheta_{\rm{\rm opt}}=0.43$).}
\end{figure}
The one-body density derived from the rotated CS solution is very different from 
the GSM density, see Fig.~\ref{GSM_vs_L2_state}. As the theory implies, the CS density is localized, and  the degree of localization increases with  
$\vartheta$ \cite{backroterror}. To compare with the GSM density, which has outgoing
asymptotics, we need to back-rotate the CS radial density.

\begin{figure}[h!] 
  \includegraphics[width=\columnwidth]{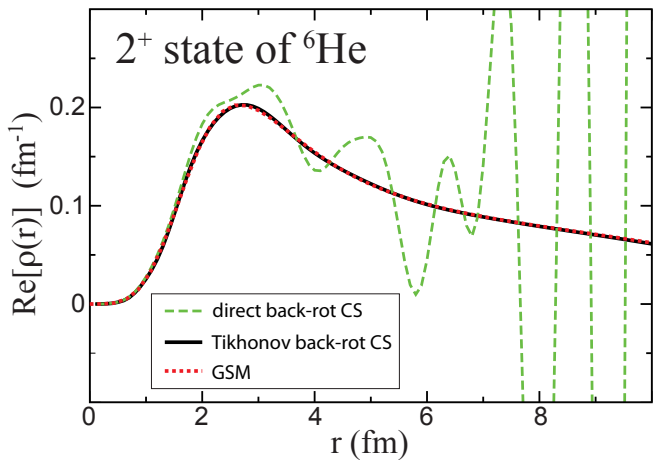}
  \caption[T]{\label{real_tikh}
  (Color online) Real part of one-neutron radial density for the unbound  2$^+$ state in $^6$He obtained in the  back-rotated
 (dashed line) and Tikhonov-regularized-back-rotated (solid)  CS-Slater method at $\vartheta_{\rm{\rm opt}}$. The GSM density is marked by a dotted line.}
\end{figure}

\begin{figure}[h!] 
  \includegraphics[width=\columnwidth]{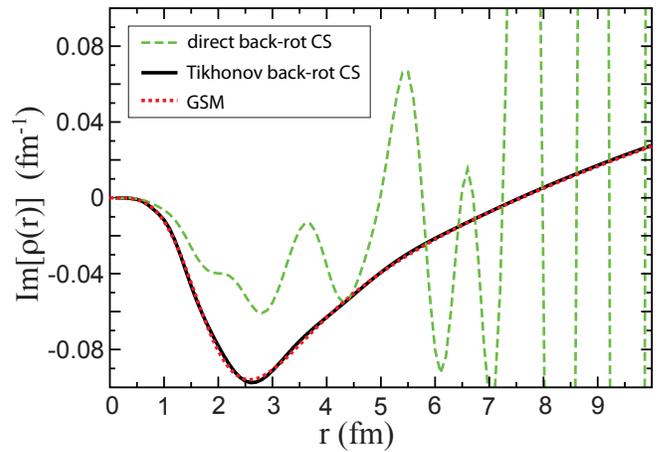}
  \caption[T]{\label{imag_tikh}
  (Color online) Similar to Fig.~\ref{real_tikh} but for the imaginary part of the density.}
\end{figure}

The comparison of the back-rotated CS-Slater and GSM 2$^+$-state densities is presented  in Figs.~\ref{real_tikh} and \ref{imag_tikh}.
Here the problem with the back-rotated CS density is far more pronounced
than for the g.s. case shown in  Fig.~\ref{den_theta}: at $r>2$\,fm, the real part of the back-rotated density exhibits unphysical oscillations. The magnitude of those oscillations grows with $\vartheta$, even if the basis size is increased. The situation is even worse for the imaginary part of the density, which does not resemble the GSM density at $r>1$\,fm. 

The numerical 
instability of the back-rotated CS wave functions is an example of an  ill-posed inverse problem~\cite{Tikh_book}. 
The amplitudes of the wave function \eqref{fdef} are determined numerically, and the associated errors 
are amplified during the  back-rotation \eqref{fdefcs} causing instabilities  seen in Figs.~\ref{real_tikh} and \ref{imag_tikh}. 
Consequently, one needs a regularization method to minimize the errors  that propagate from the coefficients to the solution.
In this paper, we apply
the Fourier analytical continuation and Tikhonov regularization procedure \cite{Tikh_orig,chu09} described in Sec.~\ref{regularization}. 

\begin{figure}[h!] 
  \includegraphics[width=\columnwidth]{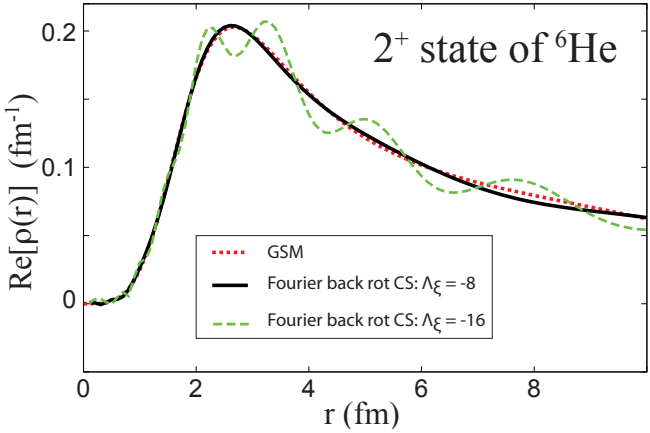}
  \caption[T]{\label{fourier_cut}
  (Color online) Real part of one-neutron radial density for the  2$^+$ resonance  in $^6$He obtained in  back-rotated CS-Slater
  using the Fourier-regularized analytical continuation with $\Lambda_{\xi}=-8$ (solid line) and $\Lambda_{\xi}=-16$ (dashed line). The GSM density is marked by a dotted line.}
\end{figure}

We first investigate the  Fourier transform \eqref{fourier_analytic_cont}, which provides us with an analytical continuation of the density.
It is understood that if one performs the integral in the full interval $(-\infty,\infty)$, the analytically-continued density would also
exhibit unwanted oscillations. Indeed, at large negative values of $\xi$ in \eqref{fourier_analytic_cont}, the exponent may become
very large amplifying  numerical errors of the Fourier transform  $\hat f_{\vartheta}(\xi)$ and causing 
numerical instabilities. For this reason we cut the lower range  of  $\xi$ in
\eqref{fourier_analytic_cont} to obtain the expression for the Fourier-regularized function:
\begin{equation}\label{fourier_analytic_cont_cut}
f_\vartheta(x+iy)=\frac{1}{\sqrt{2\pi}}\int_{\Lambda_{\xi}}^\infty e^{i(x+iy)\xi}\hat f_\vartheta (\xi) d\xi.
\end{equation}
Figure~\ref{fourier_cut} compares the GSM density of  the 2$^+$  resonance in $^6$He with back-rotated CS-Slater densities using the Fourier-regularized analytical continuation.
By taking the cutoff parameter $\Lambda_{\xi}=-8$  we obtain a density that is
almost identical to that of the GSM. With $\Lambda_{\xi}=-16$, 
the analytically-continued density starts to oscillate around the GSM result, and with even larger negative values of cutoff  used, the high-frequency components become amplified and eventually one recoups the highly-fluctuating direct back-rotation result of Fig.~\ref{real_tikh}.

In the Tikhonov method, the sharp cutoff $\Lambda_{\xi}$ is replaced by a smooth cutoff (or filter) characterized by a smoothing  parameter $\kappa$. In Eq.~\eqref{tikh_formula} this has been achieved by means of the damping function (regulator) $[1+\kappa\exp(-2y\xi)]^{-1}$ that attenuates  large negative values of $\xi$, with the parameter $\kappa$ controlling the degree of regularization. The functional form of the regulator is not
unique; it depends on the nature of the problem. In the applications presented in this study,  the analytically-continued density coincides with the  $\vartheta$-independent result for $\kappa$ = 4$\times$10$^{-4}$, which also corresponds to the original resonant GSM solution. The results presented in Figs.~\ref{real_tikh} and \ref{imag_tikh} demonstrate that both real and imaginary parts of the resonance's density obtained in the Tikhonov-regularized CS-Slater method are in excellent agreement with the GSM result.

\begin{figure}[h!] 
  \includegraphics[width=\columnwidth]{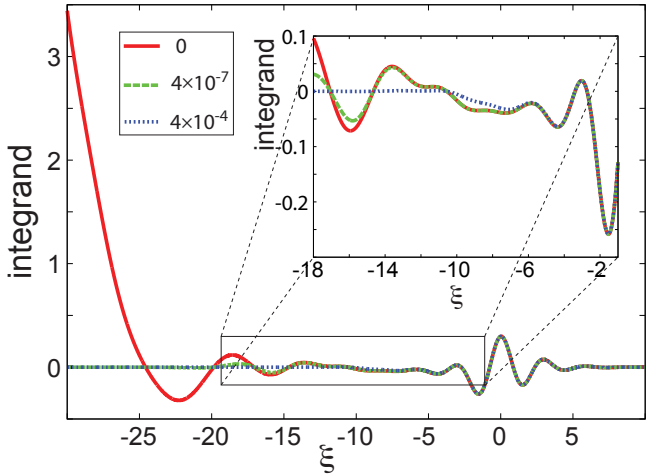}
  \caption[T]{\label{Integrand_tikhonov}
  (Color online) The real part of the integrand in Eq.~\eqref{tikh_formula}, calculated at 
  $r = 20$\,fm, $\vartheta_{\rm opt}=0.43$, and  $\kappa$=0, 4$\times$10$^{-7}$, and  4$\times$10$^{-4}$. To see the detailed behavior at small negative values of $\xi$, 
  the region of $-18\le$ $\xi$ $\le -1$ is shown in the inset.}
\end{figure}
To understand in more detail the mechanism behind the Tikhonov regularization, in Fig.\ref{Integrand_tikhonov} we display the real part of the integrand in \eqref{fourier_analytic_cont} at   $r = 20$\,fm, $\vartheta_{\rm opt}=0.43$, and  $\kappa=0$ (no regularization), $\kappa=4\times$$10^{-7}$ and  4$\times$10$^{-4}$.
In the absence of regulator, at  $\xi<-10$ the integrand exhibits oscillations with increasing amplitude.  Below $\xi = -8$, all three variants of calculations are very close; this explains the excellent agreement between GSM and back-rotated CS result
in Fig.~\ref{fourier_cut} with $\Lambda_{\xi}=-8$.  
In short, with the Tikhonov method, large values of the integrand at large negative values of $\xi$ are suppressed, thus enabling us 
to obtain an excellent reproduction of the resonant density in GSM.

\begin{figure}[h!] 
  \includegraphics[width=\columnwidth]{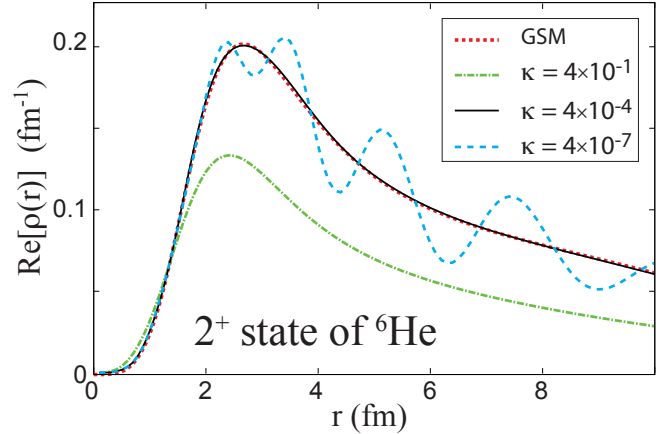}
  \caption[T]{\label{Tikhonov_method}
  (Color online) 
  Real part of one-neutron radial density for the  2$^+$ resonance  in $^6$He obtained in  back-rotated CS-Slater method using  the Tikhonov regularization with several values of smoothing parameter $\kappa$. }
\end{figure}

It is instructive to study the behavior of the analytically continued back-rotated CS density for different  Tikhonov regularization parameters $\kappa$.
As mentioned earlier, the value $\kappa$ = 4$\times$10$^{-4}$ was found to be optimal, i.e., it produces the CS-Slater densities that are closest to those of GSM. As seen in Fig.~\ref{Tikhonov_method},
for smaller values of $\kappa$, the damping function is too small to eliminate the  oscillations at large negative $\xi$ values. This is also depicted in Fig.~\ref{Integrand_tikhonov}, where for $\kappa$ = 4$\times$10$^{-7}$
unwanted oscillations of the integrand appear around $\xi \sim 16$. 
For larger values of $\kappa$, the integral is over-regulated and produces a  suppressed density profile.
Similar patterns of $\kappa$-dependence have been  found in other  studies \cite{tikh_param1,tikh_param2,tikh_param3,tikh_param4}. 
\begin{figure}[h!] 
  \includegraphics[width=\columnwidth]{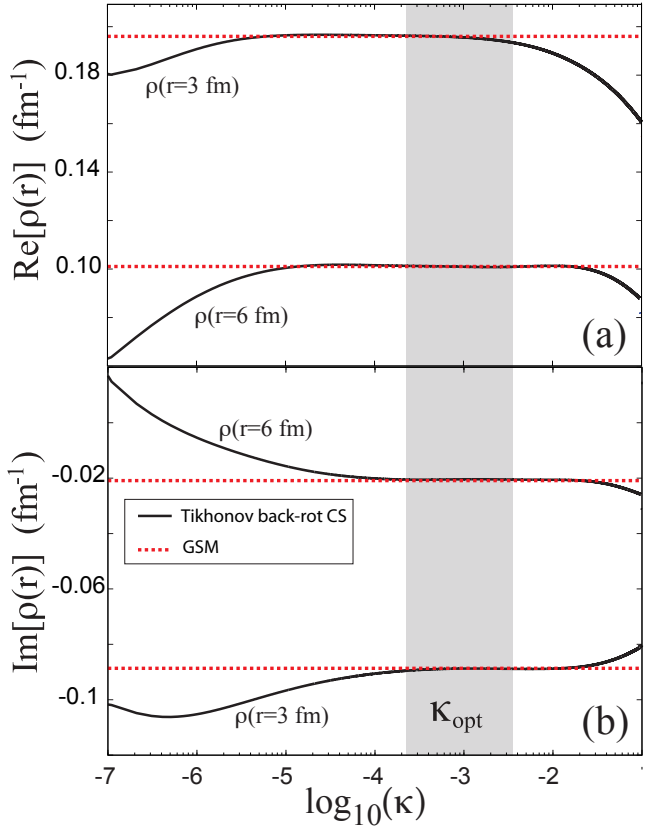}
  \caption[T]{\label{kappa_plateau}
  (Color online) 
  Real (a) and imaginary (b) parts of one-neutron radial density at  $r  =  3$ and 6 \,fm for the  $2^+$ resonance  in $^6$He, as a function
  of the Tikhonov regularization parameter $\kappa$.  In an intermediate region of $\kappa$ values (grey shading), plateaus appear that coincide with the GSM results.}
\end{figure}

The behavior seen in Fig.~\ref{Tikhonov_method} suggests a way to determine the optimal value of the smoothing parameter $\kappa$, regardless of the availability of the GSM result. The idea behind our method is presented in  Fig.~\ref{kappa_plateau}
that shows the values of $\rho(r)$ at two chosen large distances $r_\kappa$ (here $r_\kappa=3$ and 6\,fm) versus  $\kappa$ in a fairly broad range. As expected, at large and small values of $\kappa$, $\rho(r_\kappa)$ shows strong variations with the Tikhonov smoothing parameter. However,  at intermediate values,  plateau  in $\rho(r_\kappa)$ appears  that nicely coincides with the GSM results. Our optimal choice, $\kappa_{\rm opt} = 4\times10^{-4}$, belongs to  this plateau. 

\subsection{Two-body angular densities}

The two-body density contains information about two-neutron correlations. It is defined as \cite{cor_def,*cor_def2,*cor_def3}:
\begin{equation}\label{formal_def}
\rho(\bm{r},\bm{r^{\prime}}) = \langle \Psi | \delta(\bm{r}-\bm{r}_1)\delta(\bm{r^{\prime}} - \bm{r}_2) | \Psi \rangle.
\end{equation}
In spherical coordinates, it is convenient to introduce \cite{gsm_radii}
\begin{equation}\label{cor_den}
\rho(r,r^{\prime},\theta) = \langle \Psi |\delta(r_1-r)\delta(r_2-r^{\prime})\delta(\theta_{12}-\theta)|\Psi\rangle,
\end{equation}
with $r_1$ ($r_2$) being the distance between the core and the
first (second) neutron and $\theta_{12}$ - the opening angle between the
two neutrons. The density $\rho(r,r^{\prime},\theta)$
differs from the two-particle density  \eqref{formal_def} by the absence of the
Jacobian $8\pi^2 r^2 r'^2 \sin\theta$. Consequently, the two-body density is normalized according to
\begin{equation}
\int\rho(r,r^{\prime},\theta) drdr'd\theta = 1.
\end{equation}
In practical applications, \eqref{cor_den} is calculated and plotted for $r = r^{\prime}$.

By parametrizing the wave function in terms of the distance $r$ from the core nucleus and the angle $\theta$ between the valence particles,
one is able to investigate the particle correlations in the  halo nucleus.  Such calculations were performed recently \cite{gsm_radii}
to explain the observed charge radii differences in helium halo nuclei \cite{laser_prob}.
\begin{figure}[h!] 
  \includegraphics[width=\columnwidth]{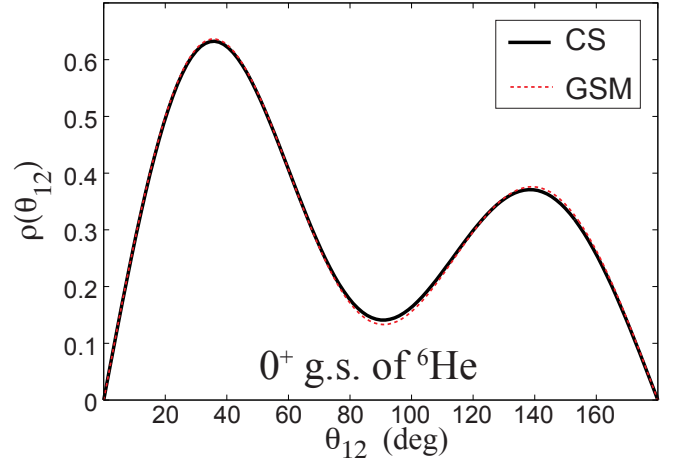}
  \caption[T]{\label{theta_den_gs}
  (Color online) Angular two-neutron density for the $^6$He g.s. predicted in  GSM and CS-Slater. }
\end{figure}
\begin{figure}[h!] 
  \includegraphics[width=\columnwidth]{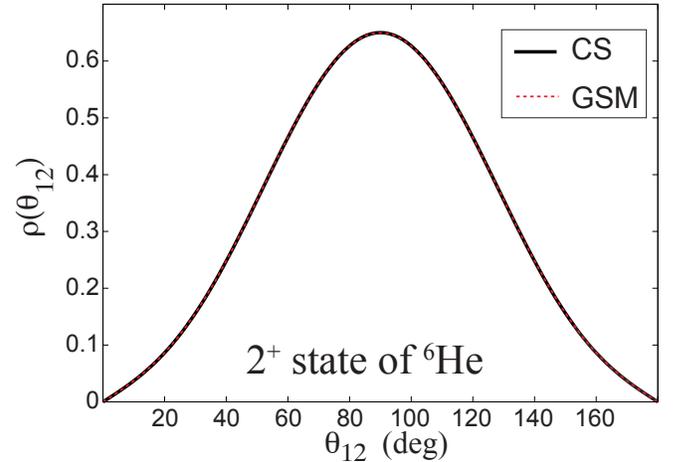}
  \caption[T]{\label{theta_den_exc}
  (Color online) Similar as in Fig.\ref{theta_den_gs} but for the 2$^+$ resonance. }
\end{figure}
To study angular correlations between valence particles, we introduce the angular density: 
\begin{equation}\label{ang_den}
\rho(\theta_{12}) = \int \, dr \int \, dr' \rho(r,r',\theta_{12}).
\end{equation}
Figures~\ref{theta_den_gs} and \ref{theta_den_exc} display $\rho(\theta_{12})$
for the g.s. and $2^+_1$ resonance, respectively.
The agreement between GSM and CS-Slater  is excellent.
It is worth noting that 
the calculation of the angular density in the CS-Slater framework does not require back-rotation. Indeed, since the CS operator \eqref{complex_rot1}  acts only on the radial coordinates, once
they are integrated out one obtains the unscaled result.

\section{Conclusions}\label{concl}

In this work, we introduce the new efficient CS method in a Slater basis to treat open many-body  systems. We apply the new technique  to  the two-neutron halo nucleus $^6$He considered as a three body problem. The interaction between valence neutrons is modelled by a finite-range Minnesota force.

To benchmark the new method, we computed the weakly bound g.s. and  $2^+_1$ resonance in $^6$He in both CS-Slater and GSM.
We carefully studied the numerical accuracy of both methods and found it more than sufficient for the purpose of benchmarking. Based on our tests,
we find  both approaches equivalent for all the quantities studied.
In a parallel development \cite{masui1,masui},  the CS method in a Gaussian basis \cite{Hiyama}
has been compared with GSM for $^6$He and $^6$Be and a good overall agreement has been found.

An important aspect of our study was the application of the
Tikhonov regularization technique to CS-Slater back-rotated wave functions in order to minimize the  ultraviolet numerical noise  at finite scaling angles $\vartheta$. We traced back the origin of high-frequency oscillations to the high-frequency part of the Fourier transform associated with the analytic continuation of the CS wave function and found the practical way to determine the smoothing  parameter defining the Tikhonov regularization.
The applied  stabilization method  allows to reconstruct  
the correct radial asymptotic behavior by using  localized complex-scaled  wave functions. This can be of  importance when 
calculating observables  that are directly related to the asymptotic behavior of the system, such as cross sections or decay widths. The proposed method is valid not only for narrow resonances (as for example Ref.~\cite{pade}), but also for broad resonant states, such as the
 excited 2$^+$ state of $^6$He.

In the near future, we intend to include the inter-nucleon distance $r_{12}$ in Eq.~\eqref{radform} to obtain the full Hylleraas basis that promises somehow
improved numerical convergence and higher accuracy. 
This will enable us to formulate the reaction theory directly in Hylleraas coordinates. The near-term application could include
the $\alpha + d$ elastic scattering and the radiative capture reactions 
as in
\cite{katoalphad}, and atomic applications such as electron-hydrogen scattering.

\appendix*

\section{Radial integrals}

To simplify the radial integral (\ref{radint})
we use the explicit form of the Legendre polynomial
$P_\lambda(x)=\sum_{n=0}^\lambda \eta_{\lambda,n}x^n$
and  the binomial theorem to get: 
\begin{widetext}
\begin{eqnarray}\label{I1}
&I^{(\lambda)}(n_{13},n_{23})=\sum_{n=0}^\lambda {\eta_{\lambda,n}}{2^{-n}}\sum_{k=0}^n{{n}\choose{k}}\sum_{m=0}^k (-1)^k {{k}\choose{m}}\nonumber\\
&\times\int_0^\infty  dr_{13}\,r_{13}^{n_{13}+n-2k+1} e^{-a_{13}r_{13}}
\int_0^\infty dr_{23} \, r_{23}^{n_{23}+2k-2m-n+1}e^{-a_{23}r_{23}}
\int_{\vert r_{13}-r_{23}\vert}^{r_{13}+r_{23}} dr_{12}\ r_{12}^{2m+1}
V_{12}(r_{12}).
\end{eqnarray}
Now we make a variable transformation
from the relative coordinates $r_{12}, r_{13}$ and $r_{23}$  to the Hylleraas coordinates $s,t,u$  defined by the equations
$s=r_{13}+r_{23}$, 
$t=r_{13}-r_{23}$, and
$u=r_{12}$.
Expressed in $s, t$, and $u$, the radial volume element becomes 
$d\tau_r=\frac{1}{8}(s^2-t^2)ds\,du\,dt$,
and (\ref{I1}) can be written as:
\begin{eqnarray}
&I^{(\lambda)}(n_{13},n_{23})=\sum_{n=0}^\lambda \eta_{\lambda,n}{2^{-n}}\sum_{k=0}^n{{n}\choose{k}}
\sum_{m=0}^k (-1)^k {{k}\choose{m}}\sum_{k_{13}=0}^{N_{13}+1}
\sum_{k_{23}=0}^{N_{23}+1}2^{-3-N_{13}-N_{23}}\nonumber\\
&\times
{{N_{13}+1}\choose{k_{13}}}{{N_{23}+1}\choose{k_{23}}}(-1)^{k_{23}}
\int_0^\infty ds\,e^{-a s}s^{N_{13}+N_{23}+2-k_{13}-k_{23}}\int_0^s du \,u^{N_{12}} V_{12}(u)
\int_0^u dt \, 
t^{k_{13}+k_{23}}e^{-bt}, 
\end{eqnarray}
where
$a=\frac{1}{2}(a_{13}+a_{23})$, $b=\frac{1}{2}(a_{13}-a_{23})$,
$N_{12}=2m+1$, $N_{13}=n_{13}+n-2k$, and $N_{23}=n_{23}+2k-2m-n$.
With the help of the integral
\begin{equation}\label{basint}
I(n_s,n_t,n_u,a,b)=\int_0^\infty ds\,s^{n_s}e^{-as} \int_0^s du \,u^{n_u} V_{12}(u)\int_0^u dt\, t^{n_t} e^{-bt} 
\end{equation}
we can write:
\begin{eqnarray}
&I^{(\lambda)}(n_{13},n_{23})=\sum_{n=0}^\lambda {\eta_{\lambda,n}}{2^{-n}}\sum_{k=0}^n{{n}\choose{k}}
\sum_{m=0}^k (-1)^k {{k}\choose{m}}\sum_{k_{13}=0}^{N_{13}+1}\sum_{k_{23}=0}^{N_{23}+1}
{{N_{13}+1}\choose{k_{13}}}{{N_{23}+1}\choose{k_{23}}}\nonumber\\
&\times{(-1)^{k_{23}}}{2^{-3-N_{13}-N_{23}}}  I(N_{13}+N_{23}+2-k_{13}-k_{23},k_{13}+k_{23},N_{12},a,b).
\end{eqnarray}
\end{widetext}

As the integral over $t$ in \eqref{basint} can be carried out analytically and the integral over $u$ can be computed by using 
\begin{equation}
\frac{d}{ds}\left(-\frac{1}{a^{n_s+1}}\Gamma(n_s+1,as)\right)=e^{-as}s^{n_s},
\end{equation}
one gets:
\begin{eqnarray}\label{basint2}
&I(n_s,n_t,n_u,a,b) = \frac{1}{(n_t+1)a^{n_s+1}}\int_0^\infty ds\Gamma(n_s+1,as)\nonumber \\  
&\times s^{n_u+n_t+1} V_{12}(s){\rm M}(n_t+1,n_t+2,-bs), 
\end{eqnarray}
where ${\rm M}(n_t+1,n_t+2,-bs)$ is the regular confluent hypergeometric function and $\Gamma(n_s+1,as)$ is 
the incomplete Gamma function \cite{abra}. Expressing these two special functions as finite sums of elementery functions
one finally arrives at the compact general expression
\begin{eqnarray}\label{basint3}
&I(n_s,n_t,n_u,a,b)=\frac{n_s!n_t!}{a^{n_s+1}b^{n_t+1}} \times \\
&\sum_{k=0}^{n_s}\frac{a^k}{k!}\int_0^\infty ds\, e^{-as}s^{n_u+k}V(s)\left[1
- e^{-bs}\sum_{n=0}^{n_t}\frac{(bs)^n}{n!}\right],\nonumber
\end{eqnarray}
which is valid for any form factor $V(r)$. It is immediately seen that for $b=0$
\eqref{basint3} simplifies to
\begin{eqnarray}\label{basint3b0}
&I(n_s,n_t,n_u,a,0) = \frac{n_s!}{a^{n_s+1}(n_t+1)}\sum_{k=0}^{n_s}\frac{a^k}{k!}\nonumber\\
&\times\int_0^\infty ds e^{-as}V(s)s^{n_u+n_t+k+1}.
\end{eqnarray}
To compute \eqref{basint3} with  a Gaussian form factor $V(s)=\exp(-fs^2)$, we
use
\begin{equation}
\int_0^\infty ds \exp(-as-fs^2)s^n = (-1)^n K^{(n)}(a),
\end{equation}
where $K^{(n)}(z)=\frac{d^n}{dz^n}K(z)$ with
\begin{equation}
K(z)=\frac{1}{2}\sqrt {\frac{\pi}{f}}\exp\left(\frac{z^2}{4f}\right)
{\rm Erfc}\left(\frac {z}{2\sqrt f}\right).
\end{equation}
Expressed in terms  of the parabolic cylinder function  $D_{-n-1}(z)$  \cite{segura98}, $K^{(n)}(z)$ is:
\begin{eqnarray}
&K^{(n)}(z)=\frac{1}{2}\sqrt {\frac{\pi}{f}}\left(\frac{1}{2\sqrt f}\right)^n
\frac{(-1)^nn!2^{(n+1)/2}}{\sqrt \pi}\nonumber\\
&\exp(z^2/(8f))D_{-n-1}(z/\sqrt {2f}).
\end{eqnarray}

\begin{acknowledgements}
Useful discussions with G.W.F Drake and M. P{\l}oszajczak 
are gratefully acknowledged. This work was
supported by
the U.S. Department of Energy under
Contract No.\ DE-FG02-96ER40963. 
This work was supported by the T\'AMOP-4.2.2.C-11/1/KONV-2012-0001 project.
The project has been supported by the European Union, co-financed by
the European Social Fund.
An allocation of advanced computing resources was provided by the National Science Foundation. Computational resources
were provided by the National Center for Computational
Sciences (NCCS) and the National Institute for Computational
Sciences (NICS). 
\end{acknowledgements}

\bibliographystyle{apsrev4-1}
\bibliography{CS_GSM}    

\end{document}